\documentclass[fleqn,10pt]{wlscirep}
\usepackage[utf8]{inputenc}
\usepackage[T1]{fontenc}
\usepackage{adjustbox}
\usepackage{soul}
\usepackage{wrapfig}
\usepackage{epsfig}
\usepackage{graphicx}

\usepackage{amssymb}
\usepackage{amsmath}

\usepackage{xcolor}
\usepackage{amsthm}
\usepackage{mathrsfs} 
\usepackage{tabularx}
\usepackage{float}
\usepackage{times}

\newcommand{\sknj}{\left(s^{(k)}_n\right)_j} 
\newcommand{\skNj}[1]{\left(s^{(k)}_{#1}\right)_j}

\twocolumn

\newcommand{\shknj}{\left(\widehat{\mathbf{s}}^{(k)}_n\right)_j}
\newcommand{\shkNj}[1]{\left(\widehat{\mathbf{s}}^{(k)}_{#1}\right)_j}

\newcommand{\shBn}{\left(\widehat{\underline{\mathbf{s}}}^{(k)}\right)_j}
\newcommand{\shBkn}{\left(\widehat{\underline{\mathbf{s}}}\right)_j}

\newcommand{\STj}{\left(\mathbf{S}_T\right)_j}
\newcommand{\SWj}{\left(\mathbf{S}_W\right)_j}

\newcommand{\vw}{\mathbf{w}}
\newcommand{\vb}{\mathbf{b}}

\newcommand{\stknj}{\left(\widetilde{\mathbf{s}}^{(k)}_n\right)_j}

\title{Transport-based morphometry of nuclear structures of digital pathology images in cancer}

\author[1,2,*]{Mohammad Shifat-E-Rabbi}
\author[1,3]{Natasha Ironside}
\author[4]{John A Ozolek}
\author[5]{Rajendra Singh}
\author[6]{Liron Pantanowitz}
\author[1,2,7]{Gustavo K Rohde}
\affil[1]{Imaging and Data Science Laboratory, Charlottesville, VA, USA}
\affil[2]{Department of Biomedical Engineering, University of Virginia, Charlottesville, VA, USA}
\affil[3]{Department of Neurological Surgery, University of Virginia School of Medicine, Charlottesville, VA, USA}
\affil[4]{Department of Pathology, Anatomy, and Laboratory Medicine, West Virginia University, Morgantown, WV, USA}
\affil[5]{Advantagecare Physicians, New Hyde Park, NY, USA }
\affil[6]{Department of Pathology, University of Michigan Medical School, Ann Arbor, MI, USA}
\affil[7]{Department of Electrical \& Computer Engineering, University of Virginia, Charlottesville, VA, USA}

\affil[*]{mr2kz@virginia.edu}

\begin{abstract}

Alterations in nuclear morphology are useful adjuncts and even diagnostic tools used by pathologists in the diagnosis and grading of many tumors, particularly malignant tumors. Large datasets such as TCGA and the Human Protein Atlas, in combination with emerging machine learning and statistical modeling methods, such as feature extraction and deep learning techniques, can be used to extract meaningful knowledge from images of nuclei, particularly from cancerous tumors. Here we describe a new technique based on the mathematics of optimal transport for modeling the information content related to nuclear chromatin structure directly from imaging data. In contrast to other techniques, our method represents the entire information content of each nucleus relative to a template nucleus using a transport-based morphometry (TBM) framework. We demonstrate the model is robust to different staining patterns and imaging protocols, and can be used to discover meaningful and interpretable information within and across datasets and cancer types. In particular, we demonstrate morphological differences capable of distinguishing nuclear features along the spectrum from benign to malignant categories of tumors across different cancer tissue types, including tumors derived from liver parenchyma, thyroid gland, lung mesothelium, and skin epithelium. We believe these proof of concept calculations demonstrate that the TBM framework can provide the quantitative measurements necessary for performing meaningful comparisons across a wide range of datasets and cancer types that can potentially enable numerous cancer studies, technologies, and clinical applications and help elevate the role of nuclear morphometry into a more quantitative science. The source codes implementing our method is available at https://github.com/rohdelab/nuclear\_morphometry.

\end{abstract}

\begin{document}
\flushbottom
\maketitle
%
%
\thispagestyle{empty}

\section{Introduction}
Alterations in nuclear morphology have been a staple in the pathologists’ repertoire of diagnostic tools since the inception of microscopic examination of tissue \cite{orsulic2022computational, bauer2017transformation,zink2004nuclear,beale1860examination,uhler2018nuclear,martinez2015identification,demay1996art}. The morphology of a nucleus, sometimes referred to as large structure organization \cite{dillon2006gene}, is determined by its microscopic structure and degree of chromatin condensation, both of which are closely regulated by interactions between the cell and its local microenvironment \cite{mazumder2007gold, uhler2018nuclear}. Defects in the coupling of the nucleus to the cytoskeleton are one means that alter chromosomal organization, leading to genomic instability and the transformation from a benign to a malignant cell \cite{zink2004nuclear, denais2014nuclear, pfeifer2019nuclear}. In the growing field of cytopathology, nuclear morphological parameters, which include increased nuclear size, increased nuclear-to-cytoplasmic ratio, irregularities of the nuclear membrane, and abnormalities in chromatin organization, are essential visual clues to guide pathologists in diagnosis and patient management decisions \cite{zink2004nuclear,uhler2018nuclear,beale1860examination,demay1996art, martinez2015identification,kaushal2019recent}. To date, the malignant potential of a tissue specimen based on morphology has been determined by the visual microscopic inspection of diagnostic pathologists \cite{veltri2014nuclear, fischer2020nuclear}. Recent advances in computer-aided digital pathology and computational pattern recognition methods have led to a number of successful experimental applications in cancer~detection \cite{veta2014breast}, staging \cite{doyle2008automated}, prognosis prediction \cite{yu2016predicting}, drug discovery \cite{zhou2006informatics}, and cell biology \cite{shifat2020cell,boland2001neural}. These methods have the potential to perform standardized, efficient, and automated large-scale analyses of nuclear structure, with the goal of providing a quantitative method for evaluating relationships between nuclear morphological changes and cellular discovery \cite{kaushal2019recent,guillaud2004quantitative}. 




\begin{figure*}[!hbt]
\centering
\includegraphics[width=17.6cm]{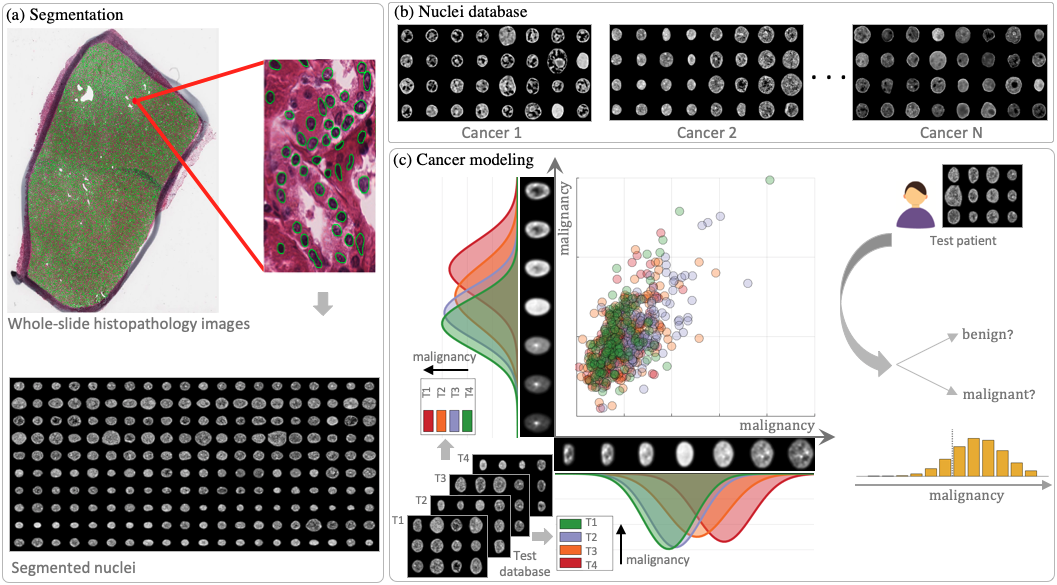}
\caption{\small{System diagram outlining the proposed cancer modeling approach. (a) Image segmentation techniques afford the ability to obtain a large-scale database of segmented nuclei from whole-slide histopathology images. (b) The proposed method takes segmented nuclei datasets obtained from various tissue types as inputs. (c) The proposed cancer modeling approach performs a joint regression in the transport space. The model can be used to visualize a specific feature, obtain malignancy potential rankings within a subset of tissue types, and classify patients, among other potential applications.}}
\label{figcm_seg}
\end{figure*}

The current practice of computational digital pathology is predominantly based on heuristic feature engineering \cite{ponomarev2014ana} and end-to-end feature learning \cite{boland2001neural}, whereby analytical features are either predetermined or learned from the data. End-to-end deep learning methods using convolutional neural networks (CNN) have obtained high classification accuracy in several experimental applications \cite{gao2016hep,qi2016exploring}, can be implemented in parallel using graphical processing units (GPUs) \cite{strigl2010performance,gu2018recent}, and are able to learn from large quantities of annotated data \cite{gu2018recent}. However, CNN methods are frequently limited by their lack of connection with an underlying physical process that can provide a scientific rationale for their use. A limited knowledge of their internal workings makes it difficult to distinguish settings when they do and do not work \cite{hosseini2017limitation}. This predisposes CNNs to unpredictably producing misleading and inaccurate results, severely restricting their interpretability and generalizability \cite{azulay2018deep,shifat2020radon}. Confidence in their results, from a mechanistic point of view, is critical for safe and effective translation to clinical or scientific use. Using these methods, it is evidently challenging to develop a model capable of quantitative nuclear morphological analysis that contributes to understanding of the relationship between the structure and function of tumor cells and the biology of cancer \cite{azulay2018deep,shifat2020cell}. Consequently, nuclear morphological studies have not yet become a quantitative science, despite being a prime target for cancer research.

While quantitative studies exist in other branches of cancer bioinformatics, including genomics and proteomics \cite{martinez2015identification,sina2018epigenetically,szuts2022fresh}, methods to build a reliable and understandable analytical model to quantify malignant transformation solely using nuclear morphological features (other than mitotic figures) have been lacking \cite{ghassemi2021false,shen2019artificial}. In addition to the aforementioned limitations of end-to-end methods, current computational systems lack robustness to adversarial information. Slight changes in image data (e.g., different pathology staining protocols)  can cause systems to make confident but erroneous predictions. \cite{azulay2018deep,antun2020instabilities,foote2022reet,liang2020generalizability,cooper2012integrated,grapov2018rise}.  This limits the accuracy of comparisons between datasets when performing meta-analyses to produce scientifically meaningful results. Previous approaches to overcome such variability and integrate information across datasets, including transfer learning \cite{weiss2016survey} and $z$-score normalization \cite{fei2021z}, have been found  to lack reliability and generalizability \cite{langnickel2021we,soltan2021limitations}, rendering their application in biomedical inference and diagnosis limited. Consequently, it has not yet been possible to describe similarities or differences across cancer types, combine datasets to enhance clinical practice and cell biology applications, (e.g., drug discovery and biomarker discovery) \cite{goossens2015cancer,kamb2007cancer}, identify organ-specific, cancer-specific, or shared malignant signatures in nuclear morphology, or test correlations between nuclear morphological signatures and gene expression, drug efficacy, or treatment response \cite{martinez2015identification,sina2018epigenetically,ahlquist2018universal}.

\begin{figure*}[!hbt]
\centering
\includegraphics[width=15.5cm]{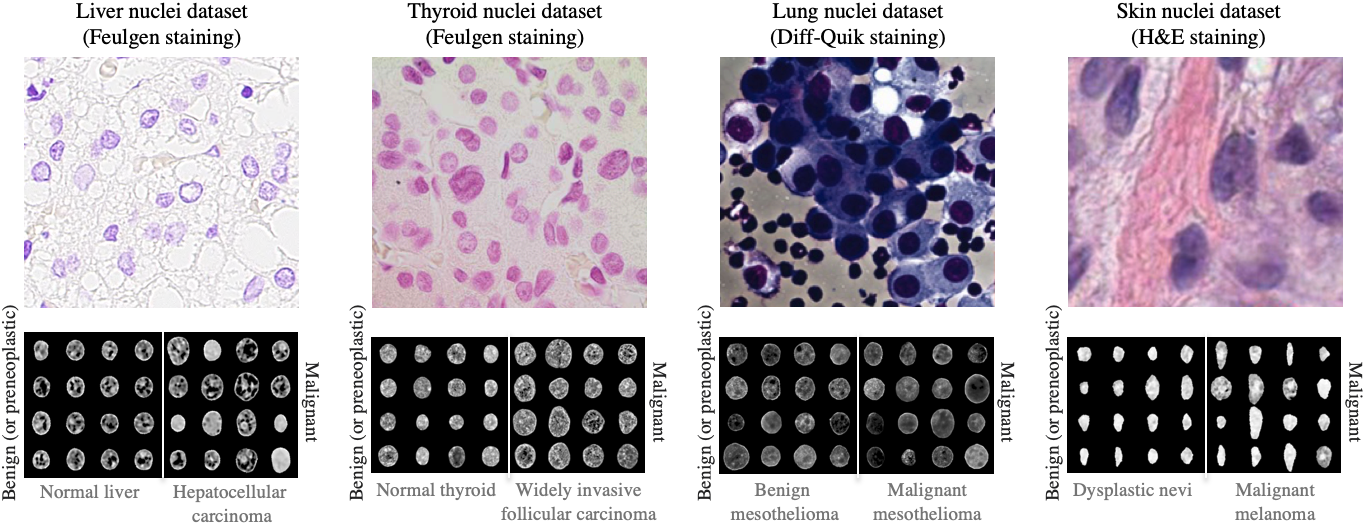}\caption{\small{Sample nuclei from digital pathology images obtained from four tissue types: liver parenchyma, thyroid gland, lung mesothelium, and skin epithelium.}}
\label{figcmf1_0}
\end{figure*}

Expanding upon earlier work \cite{wang2013linear,basu2014detecting}, we describe a mathematical technique for modeling nuclear chromatin structure and morphology directly from routinely processed and imaged tissues in the clinical setting. By considering normalized intensity values as relative measurements of chromatin density, our technique models the relative intensity observed in each pixel within a nucleus relative to a template (i.e. average) nucleus. The technique thus preserves the entire information content of each nucleus image within a biologically meaningful, transport-based, representation. Statistical analyses are then employed to summarize chromatin transport-based variabilities observed within and across datasets, as well as to elucidate meaningful discriminating information between relative malignancy levels within and across cancer types. We demonstrate our transport-based morphometry (TBM) technique can not only detect and interpret meaningful malignancy levels within each of the four cancer tissue types (liver, thyroid, lung, and skin), but also to detect and interpret persistent discriminating information along the spectrum from benign and malignant categories across these different cancer types, even when imaged using different protocols, resolutions, and staining patterns. We believe these proof of concept calculations can be used as preliminary evidence that our proposed technique can provide the quantitative measurements necessary to enable meaningful comparisons across a wide range of datasets. In combination with interesting emerging datasets (such as the human protein atlas \cite{ponten2008human}, the cancer genome atlas \cite{tomczak2015review}), we believe that our techniques can elevate the role of nuclear morphometry for use in cancer studies, technologies, and clinical applications in the emerging use of digital pathology tools to aid the pathologists, and help to render nuclear morphology studies into a more quantitative science.






\section{Problem insights}
\label{sec_plm_insight}
We begin by noting that, as microscopes measure nuclei image data from physical tissue specimens, it may be useful to consider continuum mechanics as a mathematical model for these images. During carcinogenesis, nuclei move and transform themselves according to laws often expressed in partial differential equations, such as the continuity equation \cite{kolouri2017optimal}. In other words, nuclear morphological alterations can be mathematically described as a continuous process of rearrangement of chromatin structures under the effect of biological processes. Consider the problem of modeling nuclear morphological alterations that occur during malignant transformation in a segmented nuclei image dataset (see Fig.~\ref{figcm_seg}). These morphological alterations can be the measurement of the rearrangement of chromatin structures, estimated as changes in the intensity measurements within nuclei images. 
\\\\
Given segmented nuclei image data, our goal here is to describe an approach to quantify nuclear structural changes that occur during malignant transformation that is robust to differences in staining pattern and imaging procedures. We then utilize this approach to synthesize data across multiple cancer tissue types to obtain nuclear morphological features of malignancy that are shared among cancers as a proof of concept calculation such that, like in genomics and proteomics, nuclear structure information can be used more generally than it is now. We utilize automatically segmented nuclei from histopathological images obtained from four tissue types (liver parenchyma, thyroid gland, lung mesothelium, and skin epithelium \cite{liu2016detecting}) each imaged at different resolutions and with varied staining procedures (Feulgen, Diff-Quik, and Hematoxylin and Eosin). Each dataset contained  specimens from two different histological cancer grades, which we assigned to the following classes: benign (or preneoplastic) and malignant. (see Fig.~\ref{figcmf1_0}). Details of the definitions used for each class are presented in Appendix A.

\section{Proposed approach}
Coupled with the concept of continuum mechanics mathematics to represent changes in nuclei images (as described in Section~\ref{sec_plm_insight}) and the principles of the optimal energy solution (i.e., optimal transport theory) \cite{kolouri2017optimal}, we propose an optimal mass transport-based approach to modeling nuclear morphological changes in malignancy whereby mass is represented as the image intensity \cite{kolouri2017optimal}. With the notion of a reference image (e.g., prototype nucleus), we can apply optimal transport mathematics to represent the rearrangement of the intensity measurements of nuclear chromatin structures in a physically meaningful way. Let $s(x), x \in [0,1]^2$ represent an image of a segmented nucleus, which we model as a function $s:[0,1]^2 \rightarrow \mathbb{R}_+$. As commonly assumed in transmission and fluorescence microscopy, after appropriate preprocessing (see Appendix~B) the intensity $s(x)$ is approximately proportional to the amount of mass (in our case chromatin) present at pixel location $x$ \cite{mertz2019introduction}. As the proportionality (calibration) constant is typically unknown in most routine clinical imaging procedures, we resort to normalizing it out of our analysis. That is, instead of analyzing each segmented nucleus $s(x)$ directly, given the absence of intensity calibration, we instead analyze $s(x)/\int_{[0,1]^2} s(x)dx$. Henceforth, when assume all images being analyzed have been normalized so they integrate (sum after discretization) to $1$.

Now consider two nuclear images $s_1(x), s_0(y)$, with $x,y \in [0,1]^2$. We can define the "effort" (cost) of transporting normalized intensity $s_1(x)$ from location $x$ to location $y$ as $(x-y)^2 s_0(y)$ in units of $normalized \text{ } intensity \times m^2$. Given a function that maps each coordinate from $s_0$ to $s_1$, $f(y) = x$, such that the entire normalized mass $s_0$ is transported to match $s_1$ we can define the total cost in re-arranging the normalized chromatin content from $s_1(x)$ onto $s_0(y)$ as 
\begin{equation}
    \int_{[0,1]^2} (f(y)-y)^2 s_0(y) dy
\end{equation}
where the units are once again $normalized \text{ } intensity \times m^2$. We refer to functions $f$ that re-arrange chromatin content from $s_0$ to $s_1$ as mass preserving (MP) mappings. Using the theory of optimal transport \cite{kolouri2017optimal} we can thus establish a quantitative metric (Wasserstein distance) that compares the entirety of the normalized chromatin content between two nuclear images as the solution to the following (continuous) optimization problem:
\begin{equation}
    \label{eqn:opt_cost1}W_2^2(s_0,s_1) = \inf_{f \in MP} \int_{[0,1]^2} (f(y)-y)^2 s_0(y)dy.
\end{equation}
The theory of optimal transport \cite{kolouri2017optimal} allows us to interpret and re-write the optimization problem above in terms of fluid-dynamics formulation, where we seek for velocity vector field $v(x,t)$ that transports $s_0$ onto $s_1$ by incrementally "pushing" intensities according to the continuity equation\cite{kolouri2017optimal} as follows:
\begin{align}
    W_2^2(s_0,s_1) &= \inf_{s,v} \int_0^1\int_{[0,1]^2} |v(x,t)|^2 s(x,t)dxdt,\nonumber\\
    \text{s.t.}~ & \frac{\partial s}{\partial t}+\nabla\cdot(vs)=\rho,
\end{align}
where $s(x,t)$ is the geodesic from $s_0$ to $s_1$. By solving the optimal transport continuity equation above, we can obtain the model for the rearrangement of the normalized intensity measurements from $s_0$ to $s_1$ as
\begin{align}
    s_1(x)=D_f(x)s_0(f(x)). \label{eqn_nuc_change}
\end{align}
where, $f(x)$ denotes the mass-preserving optimal transport map, and $D_f(x)$ denotes the determinant of the Jacobian matrix of $f(x)$. 
\begin{figure}[!hbt]
\centering
\includegraphics[width=8.5cm]{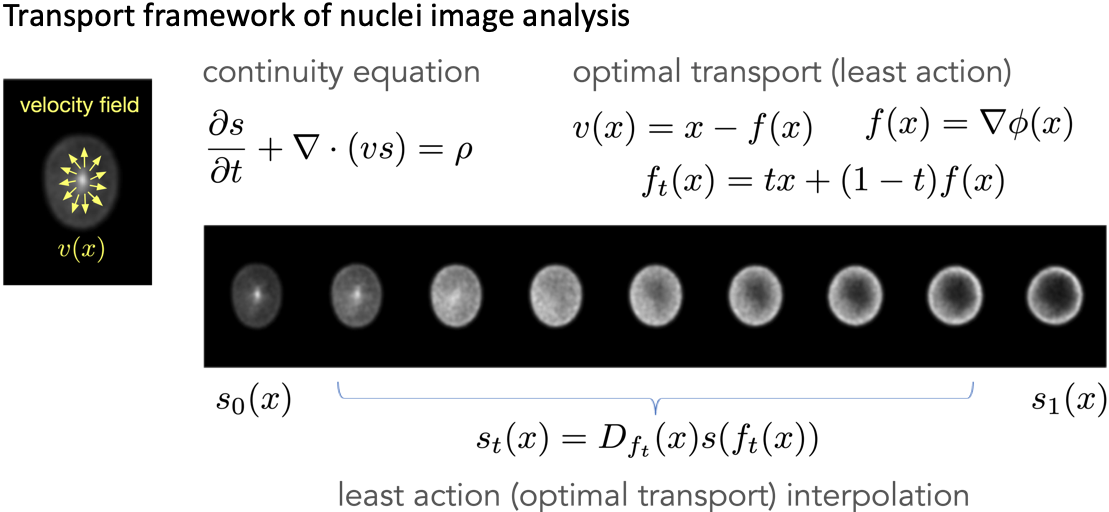}\caption{\small{Nuclei image representation using optimal transport. An image $s$ can be written in terms of a reference image $s_0$ through the use of a mapping function $f(x)$ (or equivalently, a velocity field $v(x)$). If the mapping function is chosen to be the gradient of a convex function (potential) $\phi$ then the transformation is also a metric (Wasserstein/optimal transport) between $s_0$ and the transported image $s$.}}
\label{figcmf1_1}
\end{figure}

Now, take, for example, the task of modeling the chromatin structure change between two nuclear images $s_0(x)$ and $s_1(x)$ in Figure~\ref{figcmf1_1} (leftmost and right most images). If we consider the image on the left ($s_0$) as fixed, we can represent the image on the right (or indeed any other image) by knowing $s_0$ (reference) as well as the velocity field $v(x) = x-f(x)$ that `pushes' $s_0$ forward onto some other image. Thus, we can represent any image as well as the corresponding changes in nuclear structures within that image knowing the reference $s_0$ and the map $f(x)$, or equivalently, velocity $v(x)$ field (see Figure \ref{figcmf1_1}). 

\begin{figure*}
\centering
\includegraphics[width=17.6cm]{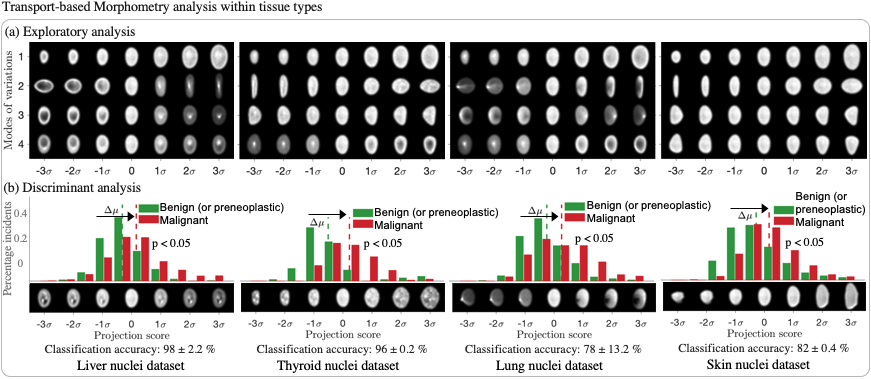}\caption{\small{The TBM framework can model nuclear morphology within a specific tissue type accurately and efficiently. The exploratory analysis shows the main trends of nuclear structural variations in the datasets. The discriminant analysis shows that the histograms of projections of the malignant class in the test set are collectively localized towards the right of the projection axis (i.e., the malignant direction obtained from the training set) with statistically significant (p < 0.05) differences in means between the benign (or preneoplastic) and the malignant classes. The discriminant analysis also demonstrates high patient classification accuracy values when obtained in the discriminant feature space.}}
\label{figcmf1}
\end{figure*}

The optimal transport-based approach has generated state of the art classification and estimation results for a variety of "segmented" signals/images including images of faces \cite{zhuang2022local}, cells \cite{shifat2020cell}, nuclei \cite{basu2014detecting}, digits \cite{rabbi2022invariance}, language characters \cite{shifat2020radon}, brain images \cite{shifat2020radon}, knee cartilage images \cite{kundu2020enabling}, ECG, physiological signals \cite{rubaiyat2022nearest,rubaiyat2022end}, and numerous other applications \cite{nishikawa2021massive,zhang2022real}. Expanding on our previous work modeling nuclear morphological changes within a specific tissue type, this paper aims to combine transport-based image transforms, which we denote as transport-based morphometry (TBM), with a set of new statistical regression methods, to synthesize and compare nuclear morphological changes between different tissue types. Our aim is to enable meaningful comparisons across a wide range of datasets, and to identify nuclear features that are shared by different cancer types. More details regarding the TBM methodology are described in Appendix B. Next, we will highlight a few important aspects of TBM, which will help us understand its role in quantitative nuclear morphometry.


\subsection{TBM provides standardized measurements}
The TBM formulation provides us a physically meaningful metric (distance), which can be used to compare two nuclear images. We can use the Wasserstein metric (see equation~\eqref{eqn:opt_cost1}) to compare two nuclear images $s_0$ and $s_1$ and quantify the relative intensity changes between them as a representation of changes in chromatin structures. Note that, the metric $W_2$ can be expressed in terms of a well-defined unit: the unit of $normalized \text{ } intensity \times m^2$. Thus, the TBM framework provides a standardized quantitative measurement of the distributions of chromatin structures where chromatin measurements are represented by the pixel intensities in nuclear images.

Because the proposed TBM approach enables us to measure the change of chromatin structures in terms of a well-defined unit, we can perform meaningful comparisons across a wide range of datasets, even when imaged using different protocols, resolutions, and staining patterns. Any nuclear morphological feature computed using the proposed approach can be expressed in terms of the same measurement unit of $normalized\text{ }intensity \times m^2$. On the contrary, the other approaches, including feature engineering and end-to-end feature learning, do not usually provide well-defined units for the computed features, which makes it challenging to perform meaningful analyses in the joint feature space, compare information across datasets, and describe similarities or differences across datasets imaged with different pathology staining protocols.



\subsection{TBM enhances interpretability}
Unlike most deep learning and feature-based approaches, the TBM formulation allows us to visualize the morphological changes between two images. The TBM framework provides a geodesic (interpolation) between images that can improve understanding of related phenomena. Take, for example, the task of filling the gap (interpolating) between two nuclear chromatin measurements $s_0(x)$ and $s_1(x)$ (leftmost and rightmost images in Figure \ref{figcmf1_1}). In addition to the transformation of the two images described by the function $f(x)$, we can obtain the Wasserstein geodesic between $s_0(x)$ and $s_1(x)$, described by the function $f_t(x)=tx+(1-t)f(x),0\leq t\leq 1$. This enables us to visualize the intermediate nuclei between $s_0(x)$ and $s_1(x)$ in the original image space (see Figure \ref{figcmf1_1}). Thus, we can visualize the process of evolution from $s_0(x)$ to $s_1(x)$, which improves interpretability of the results and enhances our understanding of the underlying physical process. 
 
\begin{figure*}[!hbt]
\centering
\includegraphics[width=17.6cm]{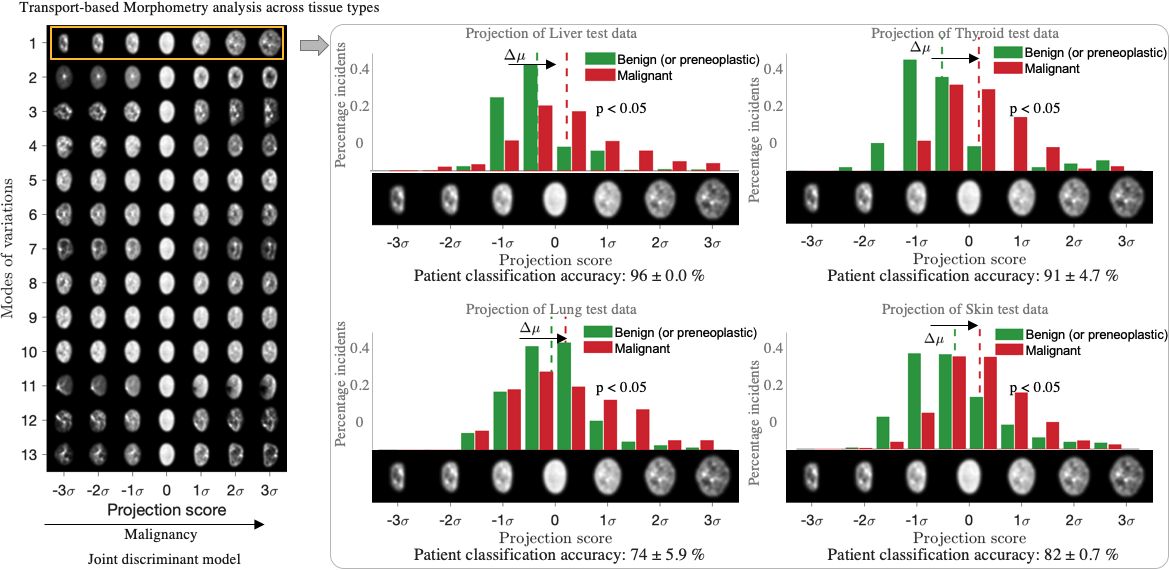}
\caption{\small{The proposed model identifies a set of nuclear morphological features of malignancy that are shared across cancer types (left panel). The projections of the test data in the malignant class of the four tissue types are collectively localized towards the right of the projection axis (i.e., the malignant direction obtained from the training set) with statistically significant ($p<0.05$) differences in means between the benign (or preneoplastic) and the malignant classes (right panel). The patient classification performances are similar to the performances of individual tissue-specific models in Fig.~\ref{figcmf1}}, indicating high discriminating capacity of the learned features.}
\label{fig:res1}
\end{figure*}


\subsection{TBM models tissue-specific morphology}
Our previous work applying TBM to tissue specimens can model nuclear morphology within a specific tissue type accurately and efficiently \cite{basu2014detecting,ozolek2014accurate,tosun2015detection,liu2016detecting,hanna2017predictive}. Fig.~\ref{figcmf1} summarizes the application of our previously described {\it{tissue-specific}} TBM model to each of the four tissue datasets we used in this study. Visual representations of the principal phenotype variability in each of the four tissue datasets (using the training set) were obtained using principal component analysis (PCA) in the transport space. \cite{basu2014detecting} We present the main trends regarding size, shape, texture, and other nuclear structural variations in each dataset (see Fig.~\ref{figcmf1}(a)). Using principal linear discriminant analysis (PLDA) in the transport space \cite{basu2014detecting}, we can visualize the principal nuclear morphological changes responsible for discriminating between the benign (or preneoplastic) and the malignant classes (see Fig.~\ref{figcmf1}(b)). In the test dataset, the histograms of the malignant class, which are obtained as projections onto the direction of principal nuclear morphological change, are collectively located towards the right (i.e., the malignant direction obtained from the training set). Differences between the histograms of the benign (or preneoplastic) and the malignant classes are statistically significant ($p<0.05$). The discriminant analysis also demonstrates high patient classification accuracy values when obtained in the discriminant feature space. Consistent with the findings of our previous papers \cite{basu2014detecting,ozolek2014accurate,tosun2015detection,liu2016detecting}, the tissue-specific TBM model can accurately model nuclear morphology within a single tissue type in each of the aforementioned datasets. This is confirmed by its meaningful visual interpretation and effective discriminatory capability in the test dataset.

\subsection{TBM for modeling shared cancer morphology}
As discussed in the previous sections, the TBM framework has been demonstrated to perform well in numerous applications, including the modeling of tissue-specific nuclear morphological changes in cancer cells. We have explained that TBM can provide a physically meaningful standardized quantitative measurement metric (i.e., the Wasserstein metric) to compare nuclear structural changes in cancer. This can be used to better understand the underlying physical mechanism that occurs during the evolution from a benign to a malignant cell. We hypothesized that the TBM framework has the potential to reliably synthesize and compare nuclear morphological information across different cancer tissue types from different datasets. In this paper, we present the results of an updated TBM framework utilizing optimal mass transport and a set of new statistical regression techniques to compare information across different datasets. More details regarding the methodology are described in Appendix B.


\section{Results}
This section demonstrates that our proposed TBM approach can synthesize data across multiple tissue types and obtain nuclear morphological features of malignancy that are shared among different cancer types. These results validate our claim that the proposed method can provide meaningful quantitative comparisons across datasets, even when imaged using different protocols, resolutions, and staining patterns. We study the effectiveness of the proposed model by evaluating 1) the projections of held-out test data on the obtained model and 2) the patient classification accuracy assessed on the held-out test data projected on the obtained model. Finally, we show an example application where our model discovers information to rank malignancy (or histological grade) within the sub-types of other unseen cancer datasets.


\begin{figure*}[!hbt]
\centering
\includegraphics[width=17.6cm]{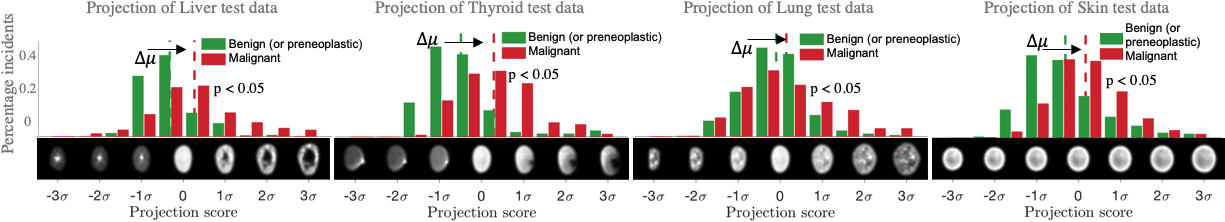}
\caption{\small{The proposed model under a modified experimental setup, where the model was trained using any three tissue types and tested using the fourth tissue type. This modified model correctly ranks the histological grade in the test tissue type using the nuclear morphological features learned from cancer tissues in the training set. The differences in projection means between the benign (or preneoplastic) and the malignant classes of the test set are statistically significant ($p<0.05$).}}
\label{figcmf3}
\end{figure*}


\subsection{Nuclear features shared across cancer types}

The proposed TBM model predicts the existence of nuclear morphological features of malignancy that are shared across cancer types and our model can identify a set of shared features. Visual representations of the identified features are shown in the left panel of Fig.~\ref{fig:res1}. Each of the identified features can be visualized as changes of nuclear images along a mode of variation, whereby changes from left to right indicate changes from the benign (or preneoplastic) to the malignant class, for each mode of variations (see Fig.~\ref{fig:res1}). The main nuclear morphological changes described by the learned features can be visualized from the nuclear images of the above visualization figure. The proposed features were obtained from the training set comprising four tissue types (liver parenchyma, thyroid gland, lung mesothelium, and skin epithelium). Fig.~\ref{fig:res1} shows the most discriminant set of nuclear morphological alterations shared between tissue types that were obtained by the model as described in Appendix B.


\subsubsection{Projection of test data on the learned feature space}




The histograms of the projections in the test set for each of the four tissue types (liver parenchyma, thyroid gland, lung mesothelium, and skin epithelium), on the proposed shared discriminant feature set are also shown in the right panel of Fig.~\ref{fig:res1}. The horizontal axis represents the spread of projections in the unit of standard deviation. The  representative image of the nuclear morphological feature corresponding to each histogram coordinate is shown below the horizontal axis. Each histogram bar indicates the percentage of nuclei in each class that closely resemble the nuclear morphological feature shown beneath that bar. 

We observe that the projections of the test data in the malignant class are collectively located toward the right (i.e., the malignant direction obtained from the training set) of the projection axis compared with the benign (or preneoplastic) class (as indicated by the location of histogram means of these two classes). This trend of collective localization of the malignant class towards the right is consistent among each of the four tissue types tested, demonstrating the shared discriminatory capability of our learned feature model. Fig.~\ref{fig:res1} shows projections on the first learned feature, however this observation can be seen in other derived shared morphological features also. The p-values of the differences of histogram-means between the benign (or preneoplastic) and the malignant classes (obtained by multivariate $t$-test) are less than $0.05$, which indicates that the separation between the two classes is statistically significant.


We further evaluated the learned model's performance using a modified experimental setup. In this experiment, we trained the model using samples from three of the four tissue types comprising liver, thyroid, lung, and skin. The model was then applied to the fourth cancer type, to predict the relative histological grade, when defined as benign (or preneoplastic) versus malignant. We repeated this experiment for all four tissue combinations and report the ranking results in Fig.~\ref{figcmf3}. The x-axis represents the spread of the projections (in the units of standard deviation) on the model trained on three cancer types. The  representative image of the nuclear morphological feature corresponding to each histogram coordinate is shown below the horizontal axis. Each histogram bar indicates the percentage of nuclei from the fourth cancer type, in each class, that closely resemble the nuclear morphological feature from the trained cancer model. In all four examples of nuclear morphological features derived from cancers affecting three different tissue types and applied to a fourth tissue type, we observed collective localization of the projections of the test data in the malignant class towards the right of the projection axis (i.e., the malignant direction obtained from the training set) with statistically significant ($p<0.05$) differences in means between the benign (or preneoplastic) and the malignant classes. This cross-validation model was observed to correctly rank the histological grade in the test tissue type using the nuclear morphological features learned from cancer tissues in the training set. It further highlights the discriminatory ability of our shared feature model. 

\begin{figure*}[!hbt]
\centering
\includegraphics[width=16.5cm]{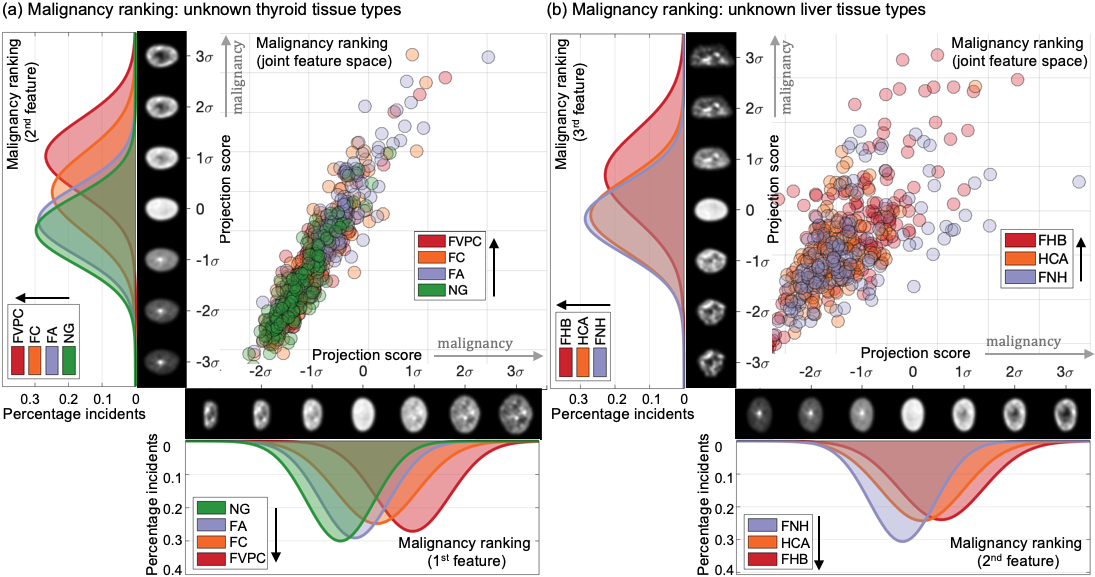}
\caption{\small{Application of the learned model in ranking the malignant potential within the subtypes of unseen cancer datasets from a particular organ: (a) malignancy ranking within the subtypes of thyroid tissue, (b) malignancy ranking within the subtypes of liver tissue. The rankings (from less malignant to more malignant) jointly predicted by the model are NG, FA, FC, FVPC and FNH, HCA, FHB for the thyroid and liver test tissue types, respectively.}}
\label{fig:res3}
\end{figure*}

\subsubsection{Patient classification using the learned features}

To evaluate the accuracy of the proposed model in ranking malignancy potential to estimate histological cancer grade, we used the learned nuclear morphological features shared between cancers affecting the four different tissue types to classify patients. We began by obtaining the histograms of projections of the nuclei for each patient in the training set on the shared discriminant morphological feature space. Next, we trained classifiers using these projections of nuclei and the corresponding histograms obtained from the training set. We obtained four sets of classifiers corresponding to the training set for each of the four tissue types. We also obtained the histograms of projections of the nuclei of the patients in the test set on the shared feature space (comprising all four tissue types). We used these test set patient histograms to test the performance of the trained classifiers in distinguishing patient samples from any of the four tissue types as benign (or preneoplastic) versus malignant. We evaluated several classifiers, including penalized discriminant analysis (PLDA), linear discriminant analysis (LDA), random forests (RF), logistic regression (LR), linear support vector machine (SVM-l), kernel support vector machine (SVM-k), and k-nearest neighbors (kNN). The best test accuracy values for the patient classification are provided in Fig.~\ref{fig:res1}. It can be seen that the proposed model provides reasonably high patient classification test accuracy as compared with the chance accuracy (50\%). The mean patient accuracy reached as high as $96\%, 91\%, 74\%,$ and $82.
\%$ for liver, thyroid, lung, and skin tissue types, respectively. These results are similar to the performances of the individual tissue-specific models presented in Fig.~\ref{figcmf1} for cancers affecting each of liver, thyroid, lung, and skin tissue types, respectively, which indicates high discriminating capacity of the shared cancer feature model when applied to heterogeneous tissue data. The detailed classification results are available in Appendix C.

\subsection{Application: Discovery of malignancy ranking within subtypes of unseen cancer datasets}


The proposed shared cancer features explained in the previous sections have the potential to be used in many clinical and scientific applications, such as cancer screening for early diagnosis, prognostication, therapeutic development, biophysical studies of disease pathology, and large database analyses. As explained before, the learned features were obtained from four datasets in the training set. Here, we show an application where we utilize these learned features to rank the malignant potential within the subtypes of other unseen cancer datasets from a particular organ (e.g., thyroid and liver).

Fig.~\ref{fig:res3} shows the nuclear morphology-based histological grade ranking results among different cancer types in the thyroid and the liver. The x and y axes represent the spread of the projections (in the units of standard deviation) on a particular feature learned by the model. The images beneath and to the left of the x and y axes, respectively, represent the most discriminant nuclear morphological feature corresponding to each histogram coordinate. The corresponding Gaussian curves represent the mean and standard deviation of the projections of the test cancer types on the nuclear morphological features of the trained cancer model. The scatter plots show the projections of the different cancer types for each tissue (thyroid and liver) in the joint feature space. We show the ranking results obtained from individual features as well as the results obtained jointly on the five most discriminatory nuclear morphological features learned by the shared cancer model. The rankings (from less malignant to more malignant) jointly predicted by the model for the thyroid test tissue type are nodular goiter (NG), follicular adenoma (FA), follicular carcinoma (FC), and follicular variant of papillary carcinoma (FVPC). The rankings (from less malignant to more malignant) jointly predicted by the model for the liver test tissue type is follicular nodular hyperplasia (FNH), hepatocellular adenoma (HCA), and fetal hepatoblastoma (FHB). These results demonstrate the potential for our model to rank or grade a spectrum of cancer types that vary in their malignant potential. They also highlight the out-of-distribution performance of our model.


\section{Discussion and Conclusions}
Improved understanding of the molecular mechanisms underpinning carcinogenesis has led to identification of biomarkers for risk assessment in cancer patients as we move further into the era of personalized medicine \cite{sawyers2008cancer, goossens2015cancer}. Molecular biomarkers, derived from genomic and/or proteomics means primarily, are used clinically for diagnosis, prognosis, therapeutic interventions, and following cancer progression during treatment \cite{goossens2015cancer, oldenhuis2008prognostic}. In recent years, attention has turned towards identification of biomarkers applicable across a number of cancer types \cite{arora2022universal, sina2018epigenetically}. Cross-cancer or universal cancer biomarkers may permit development of cost-effective and efficient cancer screening methods, elucidate common carcinogenesis pathways, and identify shared resistance and sensitivities to treatment \cite{levenson2010dna, arora2022universal, sina2018epigenetically}. 

For a cancer biomarker to be clinically useful, it must address a specific stage in tumor development, reliably estimate risk and be actionable \cite{tockman1992considerations, goossens2015cancer}. As noted, nuclear morphological alterations are a feature commonly utilized by pathologists to grade tumors\cite{fischer2020nuclear, fischer2010cytologic}. By estimating the degree of deviation of the nuclear appearance from a normal cell, a histological grade is assigned and utilized to inform prognostic and therapeutic decisions \cite{fischer2010cytologic, carriaga1995histologic}. Because nuclear morphological alterations affect all tumor cells, they represent a potentially useful biomarker for simultaneously evaluating multiple cancer types. 

Computational studies using feature engineering and neural network-based methods attempting to model nuclear morphological alterations have suffered from numerous drawbacks including a limited knowledge of their
internal workings which makes it difficult to safely implement them into a critical system, a requirement for large amounts of training data and a lack of robustness to adversarial information. \cite{veta2014breast,yu2016predicting,zhou2006informatics}In addition, a reliable machine learning method has not yet been found that can synthesize and summarize information across different cancers. This may be due to the lack of a quantitative metric that can be used to compare two nuclear images of any given cancer type. In this study, we present a TBM framework that uses a standardized quantitative metric (Wasserstein distance) to compare the entirety of the normalized chromatin content between two nuclear images. Our method, which preserves the information content of each image within a biologically meaningful, transport-based, representation, offers several advantages. First, it permits visualization of the change (or evolution) in nuclear structure between benign and malignant cells, enhancing understanding of the underlying biophysical process. Second, it detects and interprets persistent discriminating information between benign and malignant cells. Third, it can categorize and stratify patients or tissues by their histological grade in both known and unknown cancer types. Fourth, not only is it robust to variations in staining protocols and image resolutions, but it is also able to identify features shared by four different tissue types, enabling comparisons across a range of datasets.



In this paper, we presented visual exploratory analyses to highlight the common nuclear structural changes that were shared by cancers affecting the four tissue types. Our multivariate statistical analysis found a significant difference between the shared features discriminating benign or normal from malignant nuclei. This was consistent across all four tissue types. When we examined the discriminative ability of our shared feature model in classifying patients, we found it to estimate the tumor grade (malignancy ranking) with similar accuracy to the tissue-specific model. We further cross-validated our model's performance and found it to correctly estimate the tumor grade (malignancy ranking) in a tissue sample it was not previously trained upon, further highlighting the out of distribution performance of our shared feature model. Finally, we demonstrated our model's capacity to stratify unknown cancer subtypes acquired from a single tissue type that varied by their malignant potential. Our proposed method offers a novel approach to modeling the nuclear structure in cancer cells. We found it to accurately identify and measure the morphological changes that affected malignant cells, shared by the four different tissue types. 

Several limitations to this work must be acknowledged. Our model was derived from a small cohort of patient samples acquired from a limited number of centers and used digitized histological images from only four tissue types. In the absence of external validation, the broader generalizability of our results to a wide range of cancer types and to larger databases of patient samples remains unknown. Our analysis was solely based upon image features and we were unable to account for potential confounders including patient demographics, medical history information and treatment-related factors.

Our contributions in this paper are to (1) introduce a quantitative measurement metric that can be reliably used to discriminate nuclear morphological features of cancer cells, accounting for different tissue types, (2) demonstrate the potential for transport-based morphometry to overcome the limitations inherent to current techniques in digital pathology, and (3) present preliminary evidence that our transport-based morphometry method can make meaningful comparisons across a wide range of nuclei data. In combination with large datasets such as the human protein atlas and the cancer genome atlas, we believe that our proposed method has the potential to enable numerous clinical and scientific studies in fields such as population-based screening, development of personalized therapies, risk stratification, assessment of treatment response and understanding of carcinogenesis by, eventually, helping to elevate the potential role of nuclear morphometry as a universal cancer biomarker into a more quantitative science.


\section*{Acknowledgements}
This work was supported in part by NIH grant GM130825, NSF grant 1759802, and CSBC grant U54-CA274499.

\appendix

\section*{Appendix A: Computational Experiments}
\subsection*{Experimental setup}
The proposed model was built using a portion (training set) of the dataset, and the performance of the model was evaluated using the remaining portion (test set). Two different partitions of training and test set were selected using the training and test set splitting procedure of a 2-fold cross validation method. Then the exploratory visual and discriminant analysis (as described in the Results section) were performed using these training and test sets. The p-values of the differences of histograms between the benign (or preneoplastic) and the malignant classes were obtained by multivariate t-test. The details of patient classification procedure are explained in Appendix C.

\subsection*{Datasets}
We used four cancer pathology datasets containing the labeled nuclear microscopy images of liver, thyroid, lung, and skin cells. Tissue blocks of thyroid and liver nuclei datasets were collected from the University of Pittsburgh Medical Center \cite{wang2010optimal,basu2014detecting,ozolek2014accurate} and individual nuclei were segmented out according to \cite{wang2010optimal,basu2014detecting}. Cytology slides for lung tissue were obtained from the departments of pathology of Allegheny General Hospital and the West Penn Hospital and individual nuclei were segmented out according to \cite{tosun2015detection}. Under an Institutional Review Board approval, hematoxylin and eosin (H\&E) slides of skin tissue were retrieved from the pathology archives \cite{hanna2017predictive}. Each dataset contains a benign (or preneoplastic) population (normal liver, normal thyroid, benign mesothelioma, dysplastic nevi melanoma) and a malignant population (hepatocellular liver carcinoma, widely invasive thyroid follicular carcinoma, malignant mesothelioma, malignant melanoma). Details of the definitions used for each class are presented in Table~\ref{table:can_type}.
\begin{table*}[!hbt]
\caption{\small{Details of the definitions used for each class of different tissue types}}
\label{table:can_type}
\begin{adjustbox}{width=15cm,center}
\begin{tabular}{l|cc}
\hline
                       & Benign (or preneoplastic) & Malignant                       \\ \hline
Liver nuclei (Feulgen staining)   & Normal liver            & Hepatocellular carcinoma             \\
Thyroid nuclei (Feulgen staining) & Normal thyroid          & Widely invasive follicular carcinoma \\
Lung nuclei (Diff-Quik staining)    & Benign mesothelioma                  & Malignant mesothelioma                          \\
Skin nuclei (Hematoxylin and Eosin staining)    & Dysplastic nevi         & Malignant melanoma                          \\ \hline
\end{tabular}
\end{adjustbox}
\end{table*}


\section*{Appendix B: Methods}
In previous works, the TBM technique matches the transport of a reference image to the other images in the database \cite{basu2014detecting}. The vectors or mappings that match them are defined as the embeddings of the dataset, and one can perform statistical analysis on them. Our contribution here is to present a TBM method capable of synthesizing information obtained from multiple tissue types to find a discriminating feature model shared between pathology affecting different tissues.

\subsection*{Preprocessing}
Before we begin our analysis, we normalize the nuclei images to make them invariant of certain irrelevant variations \cite{rohde2008deformation}. Nuclei images may contain certain variations that are presumably unrelated to the cancer formation (e.g., variations in orientation, location within the field of view). Our objective is to decode variations in images other than these irrelevant ones. 

To that end, the center of mass of each image was translated to the center of view of each image, the principal axis of each image was aligned to a predetermined angle, and the images were flipped to have similar intensity weight distribution by switching co-ordinates until the images were aligned. Next, data were resized so that images from each tissue dataset (liver, thyroid, lung, and skin) have the same average area and pixel dimensions. Finally, the images were normalized such that the intensity values of all pixels in an image sum to one.

\subsection*{Image transform: linear optimal transport}
The proposed image morphometry analysis begins by producing a set of unique representations for input images, simultaneously facilitating the visualization and quantitative analysis of data. The procedure makes use of the LOT distance proposed in \cite{wang2013linear} and allows for an isometric embedding (LOT embedding \cite{wang2013linear}) of input image datasets onto the standard Euclidean space. The mathematical descriptions of the above concepts are described as follows:\\\\
\emph{The LOT distance}: The LOT distance -- which measures the optimal amount of effort required to rearrange one structure to another -- was constructed based on the tangent space approximation of the underlying Riemannian manifold representing the geometry of the datasets. Let $s=\sum_{p}a_p\delta_{\Vec{x}_p}$ be the particle representation of a sample image in the dataset and $\varphi=\sum_{q}a^0_q\delta_{\Vec{y}_q}$ be the particle representation of a reference structure. The optimal transport distance or the LOT distance between $s$ and $\varphi$ is given by
\begin{align}
   \label{eqn:lot_distance} d_{OT}^2\left(s,\varphi\right)=\min_{f\in\Pi(\mu,\mu_0)}\sum_{p}\sum_{q}\left|\Vec{x}_p-\Vec{y}_q\right|^2f_{pq}
\end{align}
subject to $f_{pq}\geq0$, $\sum_{p}f_{pq}=a^0_q$, and $\sum_{q}f_{pq}=a_p$. Here, $f$ is the map corresponding to the optimal mass transport (mass corresponds to the image intesity in this context) between $s$ and $\varphi$.\\\\
\emph{The LOT embedding:} Using the optimal transport map from equation~\eqref{eqn:lot_distance}, the linear optimal transport (LOT) embedding of the sample image $s$ is defined as follows:
\begin{align}
    \label{eqn:lot}\widehat{\mathbf{s}}=\begin{bmatrix}\frac{1}{\sqrt{a^0_1}}\sum_{p}f_{p1}\Vec{x}_p, & \frac{1}{\sqrt{a^0_2}}\sum_{p}f_{p2}\Vec{x}_p, & \cdots \end{bmatrix}^T
\end{align}

\subsubsection*{LOT embedding of nuclei images}
Let $\left\{\skNj{1},\skNj{2},\cdots\right\}$ represent the samples images of the $k$-th class of the $j$-th dataset. We computed the LOT embeddings for all the sample images, i.e., $\left\{\shkNj{1},\shkNj{2},\cdots\right\}$ using equation~\eqref{eqn:lot}. The reference structure $\varphi$ was computed using the Euclidean average of all input training images. For more details, refer to \cite{wang2013linear,basu2014detecting}.

The procedure above provides linear representations for the data, greatly facilitating the use of standard geometric data analysis techniques, such as principal component analysis (PCA), linear discriminant analysis (LDA), among others. Once we obtained the embeddings using equation~\eqref{eqn:lot}, we next performed linear statistical modeling as described in the following sections.

\subsection*{Statistical modeling}
The method we propose performs a hierarchical feature extraction with a new composite penalized linear discriminant analysis and a discriminant feature filtering. The components of the statistical model are explained as follows:
\subsubsection*{Composite penalized linear discriminant analysis}
We begin with a modified version of the PLDA technique, which we name as composite penalized linear discriminant analysis (cPLDA). The main goal of the cPLDA method is to calculate a set of discriminant directions that are shared across all datasets.

Let $\shknj$ denotes the LOT embedding of the $n$-th sample image of the $k$-th class of the $j$-th dataset. At first, we centered the dataset: we computed the average of $\shknj$ over all samples and classes of the $j$-th dataset and subtracted it from all samples in the dataset. The embeddings were standardized by removing the mean and scaling to unit variance: 
\begin{align}
    \left(\widehat{s}_n^{(k)}(i)\right)_j=\frac{\left(\widehat{s}_n^{(k)}(i)\right)_j}{\sqrt{\sum_p\sum_q\sum_r\left(\widehat{s}_p^{(q)}(i)\right)_r^2}}
\end{align}

\begin{table}
\centering
\normalsize
\caption{\normalsize{Description of symbols}}
\label{table:symbols}
\begin{tabular}{ll}
\hline
Symbols                & Description    \\ \hline
$s/\sknj$ & Any sample image / the $n$-th sample image of\\&the $k$-th class of the $j$-th dataset\\
$\shknj$& LOT embedding of $\sknj$\\
$\stknj$& Filtered LOT embedding of $\sknj$\\
$\shBkn$& Average of $\shknj$ over all samples and\\&classes of the $j$-th dataset\\
$\shBn$& Average of $\shknj$ over all samples in the\\&$k$-th class of the $j$-th dataset\\
\hline
\end{tabular}
\vspace{-1em}
\end{table}

The `total scatter matrix' for the $j$-th dataset can be computed as
\begin{align}
\STj=\sum_k\sum_n\left(\shknj-\shBkn\right)\left(\shknj-\shBkn\right)^T\nonumber
\end{align}
where $\shBkn=\frac{1}{K}\sum_{k=1}^{K}\frac{1}{N_k}\sum_{n=1}^{N_k}\shknj$ denotes the average of the entire set of LOT embeddings of the $j$-th dataset. The `within class scatter matrix' for the $j$-th cancer type is can be computed as 
\begin{align}
\SWj=\sum_k\sum_n\left(\shknj-\shBn\right)\left(\shknj-\shBn\right)^T\nonumber
\end{align}
where $\shBn=\frac{1}{N_k}\sum_{n=1}^{N_k}\shknj$ denotes the average of the $k$-th class of the $j$-th dataset. The shared discriminant direction can be obtained by maximizing the following objective function:
\begin{align}
         \label{eqn:costfun}\arg\max_{\vw}~Z(\vw)=\frac{\vw^T\left(\sum_{j}\STj\right)\vw}{\vw^T\left(\sum_{j}\SWj+\alpha\mathbf{I}\right)\vw}.
\end{align}

The optimization equation in equation~\eqref{eqn:costfun} is equivalent to the following generalized eigen-decomposition problem:
\begin{align}
         \label{eqn:cplda}\left(\left(\sum_{j}\SWj+\alpha\mathbf{I}\right)^{-1}\left(\sum_j\STj\right)\right)\vw=\lambda\vw
\end{align}
where, $\lambda=\max_{\vw}Z(\vw)$. Let $\vw^0$ is the solution of the equation~\eqref{eqn:cplda} (also equation~\eqref{eqn:costfun}), i.e., $\lambda=Z(\vw^0)$. Then we removed the feature scaling as follows:
\begin{align}
    w^0(i)=w^0(i)\sqrt{\sum_p\sum_q\sum_r\left(\widehat{s}_p^{(q)}(i)\right)_r^2}
\end{align}

Then $\vw^0$ can be regarded as the most discriminating feature, separating the cancerous and non-cancerous populations, that is shared across the datasets.

\subsubsection*{Discriminant feature filtering}
Once we obtain the most discriminant direction $\vw^0$, we filtered out the feature corresponding to this direction (i.e., the most discriminant feature) from the LOT embeddings of the datasets. Then we continued our analysis with the filtered data to determine the other discriminant directions.

Let $\widehat{\mathbf{s}}$ denotes the LOT embedding of any sample image $s$ (we are temporarily dropping the subscripts and superscripts). Now let us first consider the following representation of $\widehat{\mathbf{s}}$:
\begin{align}
    \widehat{\mathbf{s}}=\sum_{i}c_i\vb_i;~~\mbox{with, }c_i=<\vb_i,\widehat{\mathbf{s}}>\nonumber
\end{align}
where, $\vb_i$ represents the basis vector spanning the whole ambient space. 

Note that, there exist infinite choice of basis sets that spans the ambient space; here we chose a basis set such that $\vb_1=\vw^0$. Now in order to filter out components of data corresponding to $\vw^0$ or $\vb_1$, we defined the filtered representation as follows:
\begin{align}
    \widetilde{\mathbf{s}}=\sum_{i,i\neq1}c_i\vb_i;~~\mbox{with, }c_i=<\vb_i,\widehat{\mathbf{s}}>\nonumber
\end{align}

\subsubsection*{Hierarchical feature extraction}
At the end of above procedures, we obtained the most discriminant feature direction $\vw^0$ and the data $\stknj$ filtered along the direction $\vw^0$. Next, we set $\shknj=\stknj$ and repeated the procedures above from equation~\eqref{eqn:costfun} to obtain the next iteration of most discriminant feature direction and filtered data. If $\vw^0_{(i)}$ represents the most discriminant feature obtained at the $i$-th iteration, we defined the set of most discriminant universal cancer features as follows:  
\begin{align}
    \label{eqn:universal_feature}W=\{\vw^0_{(1)},\vw^0_{(2)},\vw^0_{(3)},\cdots\}
\end{align}
The effectiveness of the computed discriminant feature-set is evaluated in the experimental section. 

\section*{Appendix C}

\subsubsection*{Patient classification}

\begin{table*}[!hbt]
\caption{\small{Patient classification in the tissue-specific feature space}}
\label{table:patient_classify_single}
\begin{adjustbox}{width=18cm,center}
\begin{tabular}{ccccccccccccc}
\hline
      & \multicolumn{4}{c}{Histogram means}               & \multicolumn{4}{c}{Single nucleus classification (with maximum voting)} & \multicolumn{4}{c}{Complete histograms}                     \\
      & LV        & THY      & LNG & SKN     & LV         & THY        & LNG & SKN     & LV        & THY       & LNG & SKN     \\ \hline
LDA   & 87 $\pm$ 8.7 & 72 $\pm$ 0.9 & 59 $\pm$ 2.9 & 71 $\pm$ 3.6 & 74 $\pm$ 4.3 & 80 $\pm$ 1.3 & 74 $\pm$ 2.9  & 68 $\pm$ 0.4 & 96 $\pm$ 4.3 & 68 $\pm$ 5.1 & 69 $\pm$ 10.3 & 60 $\pm$ 5.0 \\
PLDA  & 91 $\pm$ 4.3 & 68 $\pm$ 5.1 & 72 $\pm$ 1.5 & 75 $\pm$ 1.1 & 74 $\pm$ 4.3 & 80 $\pm$ 1.3 & 74 $\pm$ 2.9  & 68 $\pm$ 0.0 & 89 $\pm$ 2.2 & 72 $\pm$ 0.9 & 72 $\pm$ 7.4  & 81 $\pm$ 1.4 \\
RF    & 91 $\pm$ 0.0 & 83 $\pm$ 3.6 & 75 $\pm$ 4.4 & 80 $\pm$ 0.7 & 91 $\pm$ 0.0 & 72 $\pm$ 0.9 & 71 $\pm$ 2.9  & 78 $\pm$ 1.4 & 93 $\pm$ 2.2 & 91 $\pm$ 0.4 & 74 $\pm$ 0.0  & 79 $\pm$ 1.4 \\
LR    & 93 $\pm$ 2.2 & 78 $\pm$ 0.9 & 69 $\pm$ 1.5 & 76 $\pm$ 2.9 & 80 $\pm$ 2.2 & 81 $\pm$ 5.7 & 72 $\pm$ 1.5  & 68 $\pm$ 0.7 & \bf{98 $\pm$ 2.2} & 70 $\pm$ 3.0 & 75 $\pm$ 7.4  & 81 $\pm$ 1.4 \\
SVM-l & 93 $\pm$ 2.2 & 66 $\pm$ 7.2 & 69 $\pm$ 4.4 & 77 $\pm$ 1.1 & 74 $\pm$ 0.0 & 24 $\pm$ 5.5 & 60 $\pm$ 10.3 & 35 $\pm$ 1.4 & 87 $\pm$ 4.3 & 67 $\pm$ 0.8 & 72 $\pm$ 7.4  & 80 $\pm$ 1.1 \\
SVM-k & 85 $\pm$ 2.2 & 76 $\pm$ 7.6 & 75 $\pm$ 4.4 & \bf{82 $\pm$ 0.4} & 89 $\pm$ 2.2 & 81 $\pm$ 5.7 & 72 $\pm$ 4.4  & 79 $\pm$ 0.4 & 87 $\pm$ 4.3 & \bf{96 $\pm$ 0.2} & 75 $\pm$ 10.3 & 81 $\pm$ 1.8 \\
kNN   & 83 $\pm$ 4.3 & 80 $\pm$ 1.3 & 71 $\pm$ 8.8 & 82 $\pm$ 0.7 & 89 $\pm$ 2.2 & 72 $\pm$ 5.3 & 71 $\pm$ 2.9  & 79 $\pm$ 0.4 & 89 $\pm$ 2.2 & 55 $\pm$ 4.5 & \bf{78 $\pm$ 13.2} & 79 $\pm$ 0.0 \\ \hline
\end{tabular}
\end{adjustbox}
\end{table*}

We began by projecting the nuclei of the patients in the training and test sets on the tissue-specific (Table~\ref{table:patient_classify_single}) or the shared (Table~\ref{table:patient_classify}) discriminant morphological feature space. Next, we trained the classifiers using three different descriptors: We used the means of histograms of projections of the nuclei, the single nucleus projections, or the complete histograms of projections of the training set to train different classifiers (see Tables~\ref{table:patient_classify_single} and \ref{table:patient_classify}). We obtained four sets of classifiers corresponding to four tissue types: liver (LV), thyroid (THY), lung (LNG), and skin (SKN). In the test phase, we used the same descriptors, i.e., the means of histograms of projections of the nuclei, the single nucleus projections, or the complete histograms of projections of the test set to predict the class of the patient. In the case of the single nucleus classification, we obtained the patient class prediction by applying the maximum voting procedure to the single nucleus class prediction results.

\begin{table*}[!hbt]
\caption{\small{Patient classification in the shared cancer feature space}}
\label{table:patient_classify}
\begin{adjustbox}{width=18cm,center}
\begin{tabular}{ccccccccccccc}
\hline
      & \multicolumn{4}{c}{Histogram means}               & \multicolumn{4}{c}{Single nucleus classification (with maximum voting)} & \multicolumn{4}{c}{Complete histograms}                     \\
      & LV        & THY      & LNG & SKN     & LV         & THY        & LNG & SKN     & LV        & THY       & LNG & SKN     \\ \hline
LDA   & 85 $\pm$ 2.2 & 76 $\pm$ 5.5 & 53 $\pm$ 5.9 & 75 $\pm$ 0.0 & 74 $\pm$ 0.0 & 79 $\pm$ 7.8 & 69 $\pm$ 4.4  & 66 $\pm$ 0.7 & 93 $\pm$ 2.2 & 80 $\pm$ 3.0  & 60 $\pm$ 1.5  & 58 $\pm$ 1.4 \\
PLDA  & 85 $\pm$ 2.2 & 83 $\pm$ 3.6 & 57 $\pm$ 1.5 & 76 $\pm$ 0.0 & 74 $\pm$ 0.0 & 79 $\pm$ 7.8 & 69 $\pm$ 4.4  & 66 $\pm$ 0.7 & 91 $\pm$ 0.0 & 76 $\pm$ 7.6  & 65 $\pm$ 0.0  & 78 $\pm$ 2.2 \\
RF    & 91 $\pm$ 4.3 & 80 $\pm$ 3.0 & \bf{74 $\pm$ 5.9} & 81 $\pm$ 0.7 & 89 $\pm$ 2.2 & 74 $\pm$ 3.2 & 69 $\pm$ 1.5  & 80 $\pm$ 0.0 & 89 $\pm$ 6.5 & \bf{91 $\pm$ 4.7}  & 69 $\pm$ 10.3 & 78 $\pm$ 0.0 \\
LR    & 87 $\pm$ 0.0 & 83 $\pm$ 3.6 & 65 $\pm$ 5.9 & 77 $\pm$ 1.1 & 80 $\pm$ 2.2 & 81 $\pm$ 5.7 & 69 $\pm$ 4.4  & 66 $\pm$ 0.0 & 91 $\pm$ 0.0 & 74 $\pm$ 1.1  & 71 $\pm$ 2.9  & 82 $\pm$ 1.1 \\
SVM-l & \bf{96 $\pm$ 0.0} & 80 $\pm$ 1.3 & 68 $\pm$ 5.9 & 79 $\pm$ 0.4 & 80 $\pm$ 6.5 & 74 $\pm$ 3.2 & 60 $\pm$ 4.4  & 37 $\pm$ 0.4 & 89 $\pm$ 2.2 & 78 $\pm$ 5.3  & 71 $\pm$ 0.0  & 78 $\pm$ 2.5 \\
SVM-k & 83 $\pm$ 0.0 & 80 $\pm$ 1.3 & 72 $\pm$ 4.4 & \bf{82 $\pm$ 0.7} & 87 $\pm$ 0.0 & 81 $\pm$ 5.7 & 72 $\pm$ 10.3 & 80 $\pm$ 0.0 & 93 $\pm$ 2.2 & 91 $\pm$ 9.1  & 69 $\pm$ 4.4  & 80 $\pm$ 1.1 \\
kNN   & 87 $\pm$ 0.0 & 83 $\pm$ 3.6 & 69 $\pm$ 1.5 & 81 $\pm$ 0.7 & 87 $\pm$ 4.3 & 68 $\pm$ 5.1 & 65 $\pm$ 8.8  & 78 $\pm$ 0.0 & 89 $\pm$ 2.2 & 61 $\pm$ 11.4 & 72 $\pm$ 1.5  & 76 $\pm$ 1.1 \\ \hline
\end{tabular}
\end{adjustbox}
\end{table*}


\bibliography{reference}

\begin{thebibliography}{10}
\urlstyle{rm}
\expandafter\ifx\csname url\endcsname\relax
  \def\url#1{\texttt{#1}}\fi
\expandafter\ifx\csname urlprefix\endcsname\relax\def\urlprefix{URL }\fi
\expandafter\ifx\csname doiprefix\endcsname\relax\def\doiprefix{DOI: }\fi
\providecommand{\bibinfo}[2]{#2}
\providecommand{\eprint}[2][]{\url{#2}}

\bibitem{orsulic2022computational}
\bibinfo{author}{Orsulic, S.}, \bibinfo{author}{John, J.},
  \bibinfo{author}{Walts, A.~E.} \& \bibinfo{author}{Gertych, A.}
\newblock \bibinfo{journal}{\bibinfo{title}{Computational pathology in ovarian
  cancer}}.
\newblock {\emph{\JournalTitle{Frontiers in Oncology}}}
  \textbf{\bibinfo{volume}{12}} (\bibinfo{year}{2022}).

\bibitem{bauer2017transformation}
\bibinfo{author}{Bauer, G.~M.} \emph{et~al.}
\newblock \bibinfo{journal}{\bibinfo{title}{The transformation of the nuclear
  nanoarchitecture in human field carcinogenesis}}.
\newblock {\emph{\JournalTitle{Future Science OA}}}
  \textbf{\bibinfo{volume}{3}}, \bibinfo{pages}{FSO206} (\bibinfo{year}{2017}).

\bibitem{zink2004nuclear}
\bibinfo{author}{Zink, D.}, \bibinfo{author}{Fischer, A.~H.} \&
  \bibinfo{author}{Nickerson, J.~A.}
\newblock \bibinfo{journal}{\bibinfo{title}{Nuclear structure in cancer
  cells}}.
\newblock {\emph{\JournalTitle{Nature reviews cancer}}}
  \textbf{\bibinfo{volume}{4}}, \bibinfo{pages}{677--687}
  (\bibinfo{year}{2004}).

\bibitem{beale1860examination}
\bibinfo{author}{Beale, L.}
\newblock \bibinfo{journal}{\bibinfo{title}{Examination of sputum from a case
  of cancer of the pharynx and the adjacent parts}}.
\newblock {\emph{\JournalTitle{Arch Med}}} \textbf{\bibinfo{volume}{2}},
  \bibinfo{pages}{1860--61} (\bibinfo{year}{1860}).

\bibitem{uhler2018nuclear}
\bibinfo{author}{Uhler, C.} \& \bibinfo{author}{Shivashankar, G.}
\newblock \bibinfo{journal}{\bibinfo{title}{Nuclear mechanopathology and cancer
  diagnosis}}.
\newblock {\emph{\JournalTitle{Trends in cancer}}}
  \textbf{\bibinfo{volume}{4}}, \bibinfo{pages}{320--331}
  (\bibinfo{year}{2018}).

\bibitem{martinez2015identification}
\bibinfo{author}{Martinez-Ledesma, E.}, \bibinfo{author}{Verhaak, R.~G.} \&
  \bibinfo{author}{Trevi{\~n}o, V.}
\newblock \bibinfo{journal}{\bibinfo{title}{Identification of a multi-cancer
  gene expression biomarker for cancer clinical outcomes using a network-based
  algorithm}}.
\newblock {\emph{\JournalTitle{Scientific reports}}}
  \textbf{\bibinfo{volume}{5}}, \bibinfo{pages}{1--14} (\bibinfo{year}{2015}).

\bibitem{demay1996art}
\bibinfo{author}{DeMay, R.~M.}
\newblock \emph{\bibinfo{title}{The art \& science of cytopathology}},
  vol.~\bibinfo{volume}{1} (\bibinfo{publisher}{Amer Society of Clinical},
  \bibinfo{year}{1996}).

\bibitem{dillon2006gene}
\bibinfo{author}{Dillon, N.}
\newblock \bibinfo{journal}{\bibinfo{title}{Gene regulation and large-scale
  chromatin organization in the nucleus}}.
\newblock {\emph{\JournalTitle{Chromosome Research}}}
  \textbf{\bibinfo{volume}{14}}, \bibinfo{pages}{117--126}
  (\bibinfo{year}{2006}).

\bibitem{mazumder2007gold}
\bibinfo{author}{Mazumder, A.} \& \bibinfo{author}{Shivashankar, G.}
\newblock \bibinfo{journal}{\bibinfo{title}{Gold-nanoparticle-assisted laser
  perturbation of chromatin assembly reveals unusual aspects of nuclear
  architecture within living cells}}.
\newblock {\emph{\JournalTitle{Biophysical journal}}}
  \textbf{\bibinfo{volume}{93}}, \bibinfo{pages}{2209--2216}
  (\bibinfo{year}{2007}).

\bibitem{denais2014nuclear}
\bibinfo{author}{Denais, C.} \& \bibinfo{author}{Lammerding, J.}
\newblock \bibinfo{journal}{\bibinfo{title}{Nuclear mechanics in cancer}}.
\newblock {\emph{\JournalTitle{Cancer biology and the nuclear envelope}}}
  \bibinfo{pages}{435--470} (\bibinfo{year}{2014}).

\bibitem{pfeifer2019nuclear}
\bibinfo{author}{Pfeifer, C.~R.}, \bibinfo{author}{Irianto, J.} \&
  \bibinfo{author}{Discher, D.~E.}
\newblock \bibinfo{title}{Nuclear mechanics and cancer cell migration}.
\newblock In \emph{\bibinfo{booktitle}{Cell Migrations: Causes and Functions}},
  \bibinfo{pages}{117--130} (\bibinfo{publisher}{Springer},
  \bibinfo{year}{2019}).

\bibitem{kaushal2019recent}
\bibinfo{author}{Kaushal, C.}, \bibinfo{author}{Bhat, S.},
  \bibinfo{author}{Koundal, D.} \& \bibinfo{author}{Singla, A.}
\newblock \bibinfo{journal}{\bibinfo{title}{Recent trends in computer assisted
  diagnosis (cad) system for breast cancer diagnosis using histopathological
  images}}.
\newblock {\emph{\JournalTitle{Irbm}}} \textbf{\bibinfo{volume}{40}},
  \bibinfo{pages}{211--227} (\bibinfo{year}{2019}).

\bibitem{veltri2014nuclear}
\bibinfo{author}{Veltri, R.~W.} \& \bibinfo{author}{Christudass, C.~S.}
\newblock \bibinfo{journal}{\bibinfo{title}{Nuclear morphometry, epigenetic
  changes, and clinical relevance in prostate cancer}}.
\newblock {\emph{\JournalTitle{Cancer Biology and the Nuclear Envelope}}}
  \bibinfo{pages}{77--99} (\bibinfo{year}{2014}).

\bibitem{fischer2020nuclear}
\bibinfo{author}{Fischer, E.~G.}
\newblock \bibinfo{journal}{\bibinfo{title}{Nuclear morphology and the biology
  of cancer cells}}.
\newblock {\emph{\JournalTitle{Acta cytologica}}}
  \textbf{\bibinfo{volume}{64}}, \bibinfo{pages}{511--519}
  (\bibinfo{year}{2020}).

\bibitem{veta2014breast}
\bibinfo{author}{Veta, M.}, \bibinfo{author}{Pluim, J.~P.},
  \bibinfo{author}{Van~Diest, P.~J.} \& \bibinfo{author}{Viergever, M.~A.}
\newblock \bibinfo{journal}{\bibinfo{title}{Breast cancer histopathology image
  analysis: A review}}.
\newblock {\emph{\JournalTitle{IEEE transactions on biomedical engineering}}}
  \textbf{\bibinfo{volume}{61}}, \bibinfo{pages}{1400--1411}
  (\bibinfo{year}{2014}).

\bibitem{doyle2008automated}
\bibinfo{author}{Doyle, S.}, \bibinfo{author}{Agner, S.},
  \bibinfo{author}{Madabhushi, A.}, \bibinfo{author}{Feldman, M.} \&
  \bibinfo{author}{Tomaszewski, J.}
\newblock \bibinfo{title}{Automated grading of breast cancer histopathology
  using spectral clustering with textural and architectural image features}.
\newblock In \emph{\bibinfo{booktitle}{2008 5th IEEE International Symposium on
  Biomedical Imaging: From Nano to Macro}}, \bibinfo{pages}{496--499}
  (\bibinfo{organization}{IEEE}, \bibinfo{year}{2008}).

\bibitem{yu2016predicting}
\bibinfo{author}{Yu, K.-H.} \emph{et~al.}
\newblock \bibinfo{journal}{\bibinfo{title}{Predicting non-small cell lung
  cancer prognosis by fully automated microscopic pathology image features}}.
\newblock {\emph{\JournalTitle{Nature communications}}}
  \textbf{\bibinfo{volume}{7}}, \bibinfo{pages}{1--10} (\bibinfo{year}{2016}).

\bibitem{zhou2006informatics}
\bibinfo{author}{Zhou, X.} \& \bibinfo{author}{Wong, S.~T.}
\newblock \bibinfo{journal}{\bibinfo{title}{Informatics challenges of
  high-throughput microscopy}}.
\newblock {\emph{\JournalTitle{IEEE Signal Processing Magazine}}}
  \textbf{\bibinfo{volume}{23}}, \bibinfo{pages}{63--72}
  (\bibinfo{year}{2006}).

\bibitem{shifat2020cell}
\bibinfo{author}{Shifat-E-Rabbi, M.}, \bibinfo{author}{Yin, X.},
  \bibinfo{author}{Fitzgerald, C.~E.} \& \bibinfo{author}{Rohde, G.~K.}
\newblock \bibinfo{journal}{\bibinfo{title}{Cell image classification: a
  comparative overview}}.
\newblock {\emph{\JournalTitle{Cytometry Part A}}}
  \textbf{\bibinfo{volume}{97}}, \bibinfo{pages}{347--362}
  (\bibinfo{year}{2020}).

\bibitem{boland2001neural}
\bibinfo{author}{Boland, M.~V.} \& \bibinfo{author}{Murphy, R.~F.}
\newblock \bibinfo{journal}{\bibinfo{title}{A neural network classifier capable
  of recognizing the patterns of all major subcellular structures in
  fluorescence microscope images of hela cells}}.
\newblock {\emph{\JournalTitle{Bioinformatics}}} \textbf{\bibinfo{volume}{17}},
  \bibinfo{pages}{1213--1223} (\bibinfo{year}{2001}).

\bibitem{guillaud2004quantitative}
\bibinfo{author}{Guillaud, M.} \emph{et~al.}
\newblock \bibinfo{journal}{\bibinfo{title}{Quantitative histopathological
  analysis of cervical intra-epithelial neoplasia sections: Methodological
  issues}}.
\newblock {\emph{\JournalTitle{Analytical Cellular Pathology}}}
  \textbf{\bibinfo{volume}{26}}, \bibinfo{pages}{31--43}
  (\bibinfo{year}{2004}).

\bibitem{ponomarev2014ana}
\bibinfo{author}{Ponomarev, G.~V.}, \bibinfo{author}{Arlazarov, V.~L.},
  \bibinfo{author}{Gelfand, M.~S.} \& \bibinfo{author}{Kazanov, M.~D.}
\newblock \bibinfo{journal}{\bibinfo{title}{Ana hep-2 cells image
  classification using number, size, shape and localization of targeted cell
  regions}}.
\newblock {\emph{\JournalTitle{Pattern Recognition}}}
  \textbf{\bibinfo{volume}{47}}, \bibinfo{pages}{2360--2366}
  (\bibinfo{year}{2014}).

\bibitem{gao2016hep}
\bibinfo{author}{Gao, Z.}, \bibinfo{author}{Wang, L.}, \bibinfo{author}{Zhou,
  L.} \& \bibinfo{author}{Zhang, J.}
\newblock \bibinfo{journal}{\bibinfo{title}{Hep-2 cell image classification
  with deep convolutional neural networks}}.
\newblock {\emph{\JournalTitle{IEEE journal of biomedical and health
  informatics}}} \textbf{\bibinfo{volume}{21}}, \bibinfo{pages}{416--428}
  (\bibinfo{year}{2016}).

\bibitem{qi2016exploring}
\bibinfo{author}{Qi, X.}, \bibinfo{author}{Zhao, G.}, \bibinfo{author}{Chen,
  J.} \& \bibinfo{author}{Pietik{\"a}inen, M.}
\newblock \bibinfo{journal}{\bibinfo{title}{Exploring illumination robust
  descriptors for human epithelial type 2 cell classification}}.
\newblock {\emph{\JournalTitle{Pattern Recognition}}}
  \textbf{\bibinfo{volume}{60}}, \bibinfo{pages}{420--429}
  (\bibinfo{year}{2016}).

\bibitem{strigl2010performance}
\bibinfo{author}{Strigl, D.}, \bibinfo{author}{Kofler, K.} \&
  \bibinfo{author}{Podlipnig, S.}
\newblock \bibinfo{title}{Performance and scalability of gpu-based
  convolutional neural networks}.
\newblock In \emph{\bibinfo{booktitle}{2010 18th Euromicro Conference on
  Parallel, Distributed and Network-based Processing}},
  \bibinfo{pages}{317--324} (\bibinfo{organization}{IEEE},
  \bibinfo{year}{2010}).

\bibitem{gu2018recent}
\bibinfo{author}{Gu, J.} \emph{et~al.}
\newblock \bibinfo{journal}{\bibinfo{title}{Recent advances in convolutional
  neural networks}}.
\newblock {\emph{\JournalTitle{Pattern Recognition}}}
  \textbf{\bibinfo{volume}{77}}, \bibinfo{pages}{354--377}
  (\bibinfo{year}{2018}).

\bibitem{hosseini2017limitation}
\bibinfo{author}{Hosseini, H.}, \bibinfo{author}{Xiao, B.},
  \bibinfo{author}{Jaiswal, M.} \& \bibinfo{author}{Poovendran, R.}
\newblock \bibinfo{title}{On the limitation of convolutional neural networks in
  recognizing negative images}.
\newblock In \emph{\bibinfo{booktitle}{2017 16th IEEE International Conference
  on Machine Learning and Applications (ICMLA)}}, \bibinfo{pages}{352--358}
  (\bibinfo{organization}{IEEE}, \bibinfo{year}{2017}).

\bibitem{azulay2018deep}
\bibinfo{author}{Azulay, A.} \& \bibinfo{author}{Weiss, Y.}
\newblock \bibinfo{journal}{\bibinfo{title}{Why do deep convolutional networks
  generalize so poorly to small image transformations?}}
\newblock {\emph{\JournalTitle{arXiv preprint arXiv:1805.12177}}}
  (\bibinfo{year}{2018}).

\bibitem{shifat2020radon}
\bibinfo{author}{Shifat-E-Rabbi, M.} \emph{et~al.}
\newblock \bibinfo{journal}{\bibinfo{title}{Radon cumulative distribution
  transform subspace modeling for image classification}}.
\newblock {\emph{\JournalTitle{Journal of Mathematical Imaging and Vision}}}
  \textbf{\bibinfo{volume}{63}}, \bibinfo{pages}{1185--1203}
  (\bibinfo{year}{2021}).

\bibitem{sina2018epigenetically}
\bibinfo{author}{Sina, A. A.~I.} \emph{et~al.}
\newblock \bibinfo{journal}{\bibinfo{title}{Epigenetically reprogrammed
  methylation landscape drives the dna self-assembly and serves as a universal
  cancer biomarker}}.
\newblock {\emph{\JournalTitle{Nature communications}}}
  \textbf{\bibinfo{volume}{9}}, \bibinfo{pages}{1--13} (\bibinfo{year}{2018}).

\bibitem{szuts2022fresh}
\bibinfo{author}{Sz{\"u}ts, D.}
\newblock \bibinfo{journal}{\bibinfo{title}{A fresh look at somatic mutations
  in cancer}}.
\newblock {\emph{\JournalTitle{Science}}} \textbf{\bibinfo{volume}{376}},
  \bibinfo{pages}{351--352} (\bibinfo{year}{2022}).

\bibitem{ghassemi2021false}
\bibinfo{author}{Ghassemi, M.}, \bibinfo{author}{Oakden-Rayner, L.} \&
  \bibinfo{author}{Beam, A.~L.}
\newblock \bibinfo{journal}{\bibinfo{title}{The false hope of current
  approaches to explainable artificial intelligence in health care}}.
\newblock {\emph{\JournalTitle{The Lancet Digital Health}}}
  \textbf{\bibinfo{volume}{3}}, \bibinfo{pages}{e745--e750}
  (\bibinfo{year}{2021}).

\bibitem{shen2019artificial}
\bibinfo{author}{Shen, J.} \emph{et~al.}
\newblock \bibinfo{journal}{\bibinfo{title}{Artificial intelligence versus
  clinicians in disease diagnosis: systematic review}}.
\newblock {\emph{\JournalTitle{JMIR medical informatics}}}
  \textbf{\bibinfo{volume}{7}}, \bibinfo{pages}{e10010} (\bibinfo{year}{2019}).

\bibitem{antun2020instabilities}
\bibinfo{author}{Antun, V.}, \bibinfo{author}{Renna, F.},
  \bibinfo{author}{Poon, C.}, \bibinfo{author}{Adcock, B.} \&
  \bibinfo{author}{Hansen, A.~C.}
\newblock \bibinfo{journal}{\bibinfo{title}{On instabilities of deep learning
  in image reconstruction and the potential costs of ai}}.
\newblock {\emph{\JournalTitle{Proceedings of the National Academy of
  Sciences}}} \textbf{\bibinfo{volume}{117}}, \bibinfo{pages}{30088--30095}
  (\bibinfo{year}{2020}).

\bibitem{foote2022reet}
\bibinfo{author}{Foote, A.}, \bibinfo{author}{Asif, A.},
  \bibinfo{author}{Rajpoot, N.} \& \bibinfo{author}{Minhas, F.}
\newblock \bibinfo{journal}{\bibinfo{title}{Reet: Robustness evaluation and
  enhancement toolbox for computational pathology}}.
\newblock {\emph{\JournalTitle{arXiv preprint arXiv:2201.12311}}}
  (\bibinfo{year}{2022}).

\bibitem{liang2020generalizability}
\bibinfo{author}{Liang, X.}, \bibinfo{author}{Nguyen, D.} \&
  \bibinfo{author}{Jiang, S.~B.}
\newblock \bibinfo{journal}{\bibinfo{title}{Generalizability issues with deep
  learning models in medicine and their potential solutions: illustrated with
  cone-beam computed tomography (cbct) to computed tomography (ct) image
  conversion}}.
\newblock {\emph{\JournalTitle{Machine Learning: Science and Technology}}}
  \textbf{\bibinfo{volume}{2}}, \bibinfo{pages}{015007} (\bibinfo{year}{2020}).

\bibitem{cooper2012integrated}
\bibinfo{author}{Cooper, L.~A.} \emph{et~al.}
\newblock \bibinfo{journal}{\bibinfo{title}{Integrated morphologic analysis for
  the identification and characterization of disease subtypes}}.
\newblock {\emph{\JournalTitle{Journal of the American Medical Informatics
  Association}}} \textbf{\bibinfo{volume}{19}}, \bibinfo{pages}{317--323}
  (\bibinfo{year}{2012}).

\bibitem{grapov2018rise}
\bibinfo{author}{Grapov, D.}, \bibinfo{author}{Fahrmann, J.},
  \bibinfo{author}{Wanichthanarak, K.} \& \bibinfo{author}{Khoomrung, S.}
\newblock \bibinfo{journal}{\bibinfo{title}{Rise of deep learning for genomic,
  proteomic, and metabolomic data integration in precision medicine}}.
\newblock {\emph{\JournalTitle{Omics: a journal of integrative biology}}}
  \textbf{\bibinfo{volume}{22}}, \bibinfo{pages}{630--636}
  (\bibinfo{year}{2018}).

\bibitem{weiss2016survey}
\bibinfo{author}{Weiss, K.}, \bibinfo{author}{Khoshgoftaar, T.~M.} \&
  \bibinfo{author}{Wang, D.}
\newblock \bibinfo{journal}{\bibinfo{title}{A survey of transfer learning}}.
\newblock {\emph{\JournalTitle{Journal of Big data}}}
  \textbf{\bibinfo{volume}{3}}, \bibinfo{pages}{1--40} (\bibinfo{year}{2016}).

\bibitem{fei2021z}
\bibinfo{author}{Fei, N.}, \bibinfo{author}{Gao, Y.}, \bibinfo{author}{Lu, Z.}
  \& \bibinfo{author}{Xiang, T.}
\newblock \bibinfo{title}{Z-score normalization, hubness, and few-shot
  learning}.
\newblock In \emph{\bibinfo{booktitle}{Proceedings of the IEEE/CVF
  International Conference on Computer Vision}}, \bibinfo{pages}{142--151}
  (\bibinfo{year}{2021}).

\bibitem{langnickel2021we}
\bibinfo{author}{Langnickel, L.} \& \bibinfo{author}{Fluck, J.}
\newblock \bibinfo{journal}{\bibinfo{title}{We are not ready yet: limitations
  of transfer learning for disease named entity recognition}}.
\newblock {\emph{\JournalTitle{bioRxiv}}}  (\bibinfo{year}{2021}).

\bibitem{soltan2021limitations}
\bibinfo{author}{Soltan, S.}, \bibinfo{author}{Khan, H.} \&
  \bibinfo{author}{Hamza, W.}
\newblock \bibinfo{title}{Limitations of knowledge distillation for zero-shot
  transfer learning}.
\newblock In \emph{\bibinfo{booktitle}{Proceedings of the Second Workshop on
  Simple and Efficient Natural Language Processing}}, \bibinfo{pages}{22--31}
  (\bibinfo{year}{2021}).

\bibitem{goossens2015cancer}
\bibinfo{author}{Goossens, N.}, \bibinfo{author}{Nakagawa, S.},
  \bibinfo{author}{Sun, X.} \& \bibinfo{author}{Hoshida, Y.}
\newblock \bibinfo{journal}{\bibinfo{title}{Cancer biomarker discovery and
  validation}}.
\newblock {\emph{\JournalTitle{Translational cancer research}}}
  \textbf{\bibinfo{volume}{4}}, \bibinfo{pages}{256} (\bibinfo{year}{2015}).

\bibitem{kamb2007cancer}
\bibinfo{author}{Kamb, A.}, \bibinfo{author}{Wee, S.} \&
  \bibinfo{author}{Lengauer, C.}
\newblock \bibinfo{journal}{\bibinfo{title}{Why is cancer drug discovery so
  difficult?}}
\newblock {\emph{\JournalTitle{Nature reviews Drug discovery}}}
  \textbf{\bibinfo{volume}{6}}, \bibinfo{pages}{115--120}
  (\bibinfo{year}{2007}).

\bibitem{ahlquist2018universal}
\bibinfo{author}{Ahlquist, D.~A.}
\newblock \bibinfo{journal}{\bibinfo{title}{Universal cancer screening:
  revolutionary, rational, and realizable}}.
\newblock {\emph{\JournalTitle{NPJ precision oncology}}}
  \textbf{\bibinfo{volume}{2}}, \bibinfo{pages}{1--5} (\bibinfo{year}{2018}).

\bibitem{wang2013linear}
\bibinfo{author}{Wang, W.}, \bibinfo{author}{Slep{\v{c}}ev, D.},
  \bibinfo{author}{Basu, S.}, \bibinfo{author}{Ozolek, J.~A.} \&
  \bibinfo{author}{Rohde, G.~K.}
\newblock \bibinfo{journal}{\bibinfo{title}{A linear optimal transportation
  framework for quantifying and visualizing variations in sets of images}}.
\newblock {\emph{\JournalTitle{International journal of computer vision}}}
  \textbf{\bibinfo{volume}{101}}, \bibinfo{pages}{254--269}
  (\bibinfo{year}{2013}).

\bibitem{basu2014detecting}
\bibinfo{author}{Basu, S.}, \bibinfo{author}{Kolouri, S.} \&
  \bibinfo{author}{Rohde, G.~K.}
\newblock \bibinfo{journal}{\bibinfo{title}{Detecting and visualizing cell
  phenotype differences from microscopy images using transport-based
  morphometry}}.
\newblock {\emph{\JournalTitle{Proceedings of the National Academy of
  Sciences}}} \textbf{\bibinfo{volume}{111}}, \bibinfo{pages}{3448--3453}
  (\bibinfo{year}{2014}).

\bibitem{ponten2008human}
\bibinfo{author}{Pont{\'e}n, F.}, \bibinfo{author}{Jirstr{\"o}m, K.} \&
  \bibinfo{author}{Uhlen, M.}
\newblock \bibinfo{journal}{\bibinfo{title}{The human protein atlas—a tool
  for pathology}}.
\newblock {\emph{\JournalTitle{The Journal of Pathology: A Journal of the
  Pathological Society of Great Britain and Ireland}}}
  \textbf{\bibinfo{volume}{216}}, \bibinfo{pages}{387--393}
  (\bibinfo{year}{2008}).

\bibitem{tomczak2015review}
\bibinfo{author}{Tomczak, K.}, \bibinfo{author}{Czerwinska, P.} \&
  \bibinfo{author}{Wiznerowicz, M.}
\newblock \bibinfo{journal}{\bibinfo{title}{Review the cancer genome atlas
  (tcga): an immeasurable source of knowledge}}.
\newblock {\emph{\JournalTitle{Contemporary Oncology}}}
  \textbf{\bibinfo{volume}{2015}}, \bibinfo{pages}{68--77}
  (\bibinfo{year}{2015}).

\bibitem{kolouri2017optimal}
\bibinfo{author}{Kolouri, S.}, \bibinfo{author}{Park, S.~R.},
  \bibinfo{author}{Thorpe, M.}, \bibinfo{author}{Slepcev, D.} \&
  \bibinfo{author}{Rohde, G.~K.}
\newblock \bibinfo{journal}{\bibinfo{title}{Optimal mass transport: Signal
  processing and machine-learning applications}}.
\newblock {\emph{\JournalTitle{IEEE signal processing magazine}}}
  \textbf{\bibinfo{volume}{34}}, \bibinfo{pages}{43--59}
  (\bibinfo{year}{2017}).

\bibitem{liu2016detecting}
\bibinfo{author}{Liu, C.}, \bibinfo{author}{Shang, F.},
  \bibinfo{author}{Ozolek, J.~A.} \& \bibinfo{author}{Rohde, G.~K.}
\newblock \bibinfo{journal}{\bibinfo{title}{Detecting and segmenting cell
  nuclei in two-dimensional microscopy images}}.
\newblock {\emph{\JournalTitle{Journal of Pathology Informatics}}}
  \textbf{\bibinfo{volume}{7}}, \bibinfo{pages}{42} (\bibinfo{year}{2016}).

\bibitem{mertz2019introduction}
\bibinfo{author}{Mertz, J.}
\newblock \emph{\bibinfo{title}{Introduction to optical microscopy}}
  (\bibinfo{publisher}{Cambridge University Press}, \bibinfo{year}{2019}).

\bibitem{zhuang2022local}
\bibinfo{author}{Zhuang, Y.} \emph{et~al.}
\newblock \bibinfo{journal}{\bibinfo{title}{Local sliced-wasserstein feature
  sets for illumination-invariant face recognition}}.
\newblock {\emph{\JournalTitle{arXiv preprint arXiv:2202.10642}}}
  (\bibinfo{year}{2022}).

\bibitem{rabbi2022invariance}
\bibinfo{author}{Rabbi, M. S.~E.} \emph{et~al.}
\newblock \bibinfo{journal}{\bibinfo{title}{Invariance encoding in
  sliced-wasserstein space for image classification with limited training
  data}}.
\newblock {\emph{\JournalTitle{arXiv preprint arXiv:2201.02980}}}
  (\bibinfo{year}{2022}).

\bibitem{kundu2020enabling}
\bibinfo{author}{Kundu, S.} \emph{et~al.}
\newblock \bibinfo{journal}{\bibinfo{title}{Enabling early detection of
  osteoarthritis from presymptomatic cartilage texture maps via transport-based
  learning}}.
\newblock {\emph{\JournalTitle{Proceedings of the National Academy of
  Sciences}}} \textbf{\bibinfo{volume}{117}}, \bibinfo{pages}{24709--24719}
  (\bibinfo{year}{2020}).

\bibitem{rubaiyat2022nearest}
\bibinfo{author}{Rubaiyat, A. H.~M.}, \bibinfo{author}{Shifat-E-Rabbi, M.},
  \bibinfo{author}{Zhuang, Y.}, \bibinfo{author}{Li, S.} \&
  \bibinfo{author}{Rohde, G.~K.}
\newblock \bibinfo{title}{Nearest subspace search in the signed cumulative
  distribution transform space for 1d signal classification}.
\newblock In \emph{\bibinfo{booktitle}{ICASSP 2022-2022 IEEE International
  Conference on Acoustics, Speech and Signal Processing (ICASSP)}},
  \bibinfo{pages}{3508--3512} (\bibinfo{organization}{IEEE},
  \bibinfo{year}{2022}).

\bibitem{rubaiyat2022end}
\bibinfo{author}{Rubaiyat, A. H.~M.} \emph{et~al.}
\newblock \bibinfo{journal}{\bibinfo{title}{End-to-end signal classification in
  signed cumulative distribution transform space}}.
\newblock {\emph{\JournalTitle{arXiv preprint arXiv:2205.00348}}}
  (\bibinfo{year}{2022}).

\bibitem{nishikawa2021massive}
\bibinfo{author}{Nishikawa, M.} \emph{et~al.}
\newblock \bibinfo{journal}{\bibinfo{title}{Massive image-based single-cell
  profiling reveals high levels of circulating platelet aggregates in patients
  with covid-19}}.
\newblock {\emph{\JournalTitle{Nature communications}}}
  \textbf{\bibinfo{volume}{12}}, \bibinfo{pages}{1--12} (\bibinfo{year}{2021}).

\bibitem{zhang2022real}
\bibinfo{author}{Zhang, C.} \emph{et~al.}
\newblock \bibinfo{journal}{\bibinfo{title}{Real-time intelligent
  classification of covid-19 and thrombosis via massive image-based analysis of
  platelet aggregates}}.
\newblock {\emph{\JournalTitle{medRxiv}}}  (\bibinfo{year}{2022}).

\bibitem{ozolek2014accurate}
\bibinfo{author}{Ozolek, J.~A.} \emph{et~al.}
\newblock \bibinfo{journal}{\bibinfo{title}{Accurate diagnosis of thyroid
  follicular lesions from nuclear morphology using supervised learning}}.
\newblock {\emph{\JournalTitle{Medical image analysis}}}
  \textbf{\bibinfo{volume}{18}}, \bibinfo{pages}{772--780}
  (\bibinfo{year}{2014}).

\bibitem{tosun2015detection}
\bibinfo{author}{Tosun, A.~B.}, \bibinfo{author}{Yergiyev, O.},
  \bibinfo{author}{Kolouri, S.}, \bibinfo{author}{Silverman, J.~F.} \&
  \bibinfo{author}{Rohde, G.~K.}
\newblock \bibinfo{journal}{\bibinfo{title}{Detection of malignant mesothelioma
  using nuclear structure of mesothelial cells in effusion cytology
  specimens}}.
\newblock {\emph{\JournalTitle{Cytometry Part A}}}
  \textbf{\bibinfo{volume}{87}}, \bibinfo{pages}{326--333}
  (\bibinfo{year}{2015}).

\bibitem{hanna2017predictive}
\bibinfo{author}{Hanna, M.~G.}, \bibinfo{author}{Liu, C.},
  \bibinfo{author}{Rohde, G.~K.} \& \bibinfo{author}{Singh, R.}
\newblock \bibinfo{journal}{\bibinfo{title}{Predictive nuclear chromatin
  characteristics of melanoma and dysplastic nevi}}.
\newblock {\emph{\JournalTitle{Journal of Pathology Informatics}}}
  \textbf{\bibinfo{volume}{8}}, \bibinfo{pages}{15} (\bibinfo{year}{2017}).

\bibitem{sawyers2008cancer}
\bibinfo{author}{Sawyers, C.~L.}
\newblock \bibinfo{journal}{\bibinfo{title}{The cancer biomarker problem}}.
\newblock {\emph{\JournalTitle{Nature}}} \textbf{\bibinfo{volume}{452}},
  \bibinfo{pages}{548--552} (\bibinfo{year}{2008}).

\bibitem{oldenhuis2008prognostic}
\bibinfo{author}{Oldenhuis, C.}, \bibinfo{author}{Oosting, S.},
  \bibinfo{author}{Gietema, J.} \& \bibinfo{author}{De~Vries, E.}
\newblock \bibinfo{journal}{\bibinfo{title}{Prognostic versus predictive value
  of biomarkers in oncology}}.
\newblock {\emph{\JournalTitle{European journal of cancer}}}
  \textbf{\bibinfo{volume}{44}}, \bibinfo{pages}{946--953}
  (\bibinfo{year}{2008}).

\bibitem{arora2022universal}
\bibinfo{author}{Arora, C.}, \bibinfo{author}{Kaur, D.} \&
  \bibinfo{author}{Raghava, G.~P.}
\newblock \bibinfo{journal}{\bibinfo{title}{Universal and cross-cancer
  prognostic biomarkers for predicting survival risk of cancer patients from
  expression profile of apoptotic pathway genes}}.
\newblock {\emph{\JournalTitle{Proteomics}}} \textbf{\bibinfo{volume}{22}},
  \bibinfo{pages}{e2000311} (\bibinfo{year}{2022}).

\bibitem{levenson2010dna}
\bibinfo{author}{Levenson, V.~V.}
\newblock \bibinfo{journal}{\bibinfo{title}{Dna methylation as a universal
  biomarker}}.
\newblock {\emph{\JournalTitle{Expert review of molecular diagnostics}}}
  \textbf{\bibinfo{volume}{10}}, \bibinfo{pages}{481--488}
  (\bibinfo{year}{2010}).

\bibitem{tockman1992considerations}
\bibinfo{author}{Tockman, M.~S.}, \bibinfo{author}{Gupta, P.~K.},
  \bibinfo{author}{Pressman, N.~J.} \& \bibinfo{author}{Mulshine, J.~L.}
\newblock \bibinfo{journal}{\bibinfo{title}{Considerations in bringing a cancer
  biomarker to clinical application}}.
\newblock {\emph{\JournalTitle{Cancer Research}}}
  \textbf{\bibinfo{volume}{52}}, \bibinfo{pages}{2711s--2718s}
  (\bibinfo{year}{1992}).

\bibitem{fischer2010cytologic}
\bibinfo{author}{Fischer, A.~H.} \emph{et~al.}
\newblock \bibinfo{journal}{\bibinfo{title}{The cytologic criteria of
  malignancy}}.
\newblock {\emph{\JournalTitle{Journal of cellular biochemistry}}}
  \textbf{\bibinfo{volume}{110}}, \bibinfo{pages}{795--811}
  (\bibinfo{year}{2010}).

\bibitem{carriaga1995histologic}
\bibinfo{author}{Carriaga, M.~T.} \& \bibinfo{author}{Henson, D.~E.}
\newblock \bibinfo{journal}{\bibinfo{title}{The histologic grading of cancer}}.
\newblock {\emph{\JournalTitle{Cancer}}} \textbf{\bibinfo{volume}{75}},
  \bibinfo{pages}{406--421} (\bibinfo{year}{1995}).

\bibitem{wang2010optimal}
\bibinfo{author}{Wang, W.} \emph{et~al.}
\newblock \bibinfo{journal}{\bibinfo{title}{An optimal transportation approach
  for nuclear structure-based pathology}}.
\newblock {\emph{\JournalTitle{IEEE transactions on medical imaging}}}
  \textbf{\bibinfo{volume}{30}}, \bibinfo{pages}{621--631}
  (\bibinfo{year}{2010}).

\bibitem{rohde2008deformation}
\bibinfo{author}{Rohde, G.~K.}, \bibinfo{author}{Ribeiro, A.~J.},
  \bibinfo{author}{Dahl, K.~N.} \& \bibinfo{author}{Murphy, R.~F.}
\newblock \bibinfo{journal}{\bibinfo{title}{{Deformation-based nuclear
  morphometry: Capturing nuclear shape variation in HeLa cells}}}.
\newblock {\emph{\JournalTitle{Cytometry Part A}}}
  \textbf{\bibinfo{volume}{73}}, \bibinfo{pages}{341--350}
  (\bibinfo{year}{2008}).

\end{thebibliography}

\end{document}